\documentclass[a4paper,11pt]{article}
\usepackage{jheppub} % for details on the use of the package, please
\usepackage[T1]{fontenc} % if needed
\usepackage{cancel}
\usepackage{hyperref}

\usepackage{amsmath}
\usepackage{relsize}

\usepackage{tikz,environ}
\usetikzlibrary{snakes}
\usetikzlibrary{decorations}
\usetikzlibrary{trees}
\usetikzlibrary{decorations.pathmorphing}
\usetikzlibrary{decorations.markings}
\usetikzlibrary{external}
\usetikzlibrary{intersections}
\usetikzlibrary{shapes,arrows}
\usetikzlibrary{arrows.meta}
\usetikzlibrary{calc}
\usetikzlibrary{shapes.misc}
\usetikzlibrary{decorations.text}
\usetikzlibrary{backgrounds}
\usetikzlibrary{fadings}
\usetikzlibrary{tikzmark,calc,arrows,shapes,decorations.pathreplacing}

\usetikzlibrary{shapes.geometric,positioning}

\tikzset{
	graviton/.style={decorate,line width=0.1mm, decoration={snake,amplitude=.3mm, segment length=0.8mm}},
	photon/.style={decorate, decoration={snake,amplitude=.4mm, segment length=2mm}, draw=red},
	scalar/.style={postaction={decorate},
	},
	massive/.style={postaction={decorate},
		line width=0.75mm,
	},
	massless/.style={postaction={decorate},
	},
	masslessWithDot/.style={postaction={decorate},
		decoration={
			markings,
			mark=at position 0.5 with {\fill circle (2pt);}}
	},
	massiveWithDot/.style={postaction={decorate},
		line width=0.5mm,
		decoration={
			markings,
			mark=at position 0.5 with {\fill circle (2pt);}}
	},
	massiveWithArrow/.style={postaction={decorate},
		line width=0.75mm,
		decoration={
			markings,
			mark=at position 0.5 with {\arrow{latex}}}
	},
	massiveLin/.style={postaction={decorate},
		double,
		thick,
		fill=white
	},
	massivePhi/.style={postaction={decorate},
		line width=0.75mm,
		dashed
	},
	masslessPhi/.style={postaction={decorate},
		dashed
	},
	unitaryCut/.style={postaction={draw,densely dashed,blue,thin},
		line width = 0.2cm,white
	},
	gluon/.style={decorate, draw=magenta,
		decoration={coil,amplitude=4pt, segment length=5pt}},
	partial ellipse/.style args={#1:#2:#3}{
		insert path={+ (#1:#3) arc (#1:#2:#3)}
	},
	cross/.style={cross out, draw=black, minimum size=2*(#1-\pgflinewidth), inner sep=0pt, outer sep=0pt},
	branchCut/.style={postaction={decorate},
		snake=zigzag,
		decoration = {snake=zigzag,segment length = 2mm, amplitude = 2mm}	
	}
	cross/.default={1pt}
}

\colorlet{mred}{black!30!red}
\colorlet{mgreen}{black!30!green}
\colorlet{mblue}{black!30!blue}
\colorlet{morange}{blue!70!red}

\newcommand{\realpartNLONoInsertionNoPhoton}{\begin{tikzpicture}[scale=1]
		\coordinate (e1) at (-1,1);
            \coordinate (e3) at (-0.55, 1);
            \coordinate (e5) at (-0.10, 1);
            \coordinate (e7) at (0.10, 1);
            \coordinate (e9) at (0.55, 1);
            \coordinate (e11) at (1, 1);

            \coordinate (e2) at (-1, -1);
            \coordinate (e4) at (-0.55, -1);
            \coordinate (e6) at (-0.10, -1);
            \coordinate (e8) at (0.10, -1);
            \coordinate (e10) at (0.55, -1);
            \coordinate (e12) at (1, -1);

            \coordinate (v1) at (-0.55, 0);
            \coordinate (v2) at (0.55, 0);
     
            \draw[massive] (e1) -- (e3) -- (e5);
            \draw[massive] (e7) -- (e9) -- (e11);
            \draw[massive] (e2)-- (e4) -- (e6);
            \draw[massive] (e8) -- (e10) -- (e12);

            \draw[line width=0.3mm, dashed, name path=rung4, blue] (0, 1.25) to (0, -1.25);
            \draw[fill=lightgray, opacity=1] (v1) ellipse (0.25 and 1.05);
            \draw[fill=lightgray, opacity=1] (v2)  ellipse (0.25 and 1.05);
\end{tikzpicture}}

\newcommand{\kmocvirtualnlonolab}{
\begin{tikzpicture}
                \coordinate (e1) at (-1,-1);	
	     	\coordinate (e2) at (-1, 1);
	     	\coordinate (e3) at ( 1, 1);
	     	\coordinate (e4) at ( 1,-1);
	     	
	     	\coordinate (v1) at (0,0);
     
	     	\draw[massive] (e1) -- (v1);
	     	\draw[massive] (e2) -- (v1);
	     	\draw[massive] (e4) -- (v1);
	     	\draw[massive] (e3) -- (v1);
	     	
            \draw[fill=lightgray, opacity=1] (0,0) circle (0.65);
            \draw[fill=white, opacity=1] (0,0) circle (0.35);
\end{tikzpicture}}

\graphicspath{{figures/}}
\usepackage[export]{adjustbox}

\usepackage{color}

\def\draftnote#1{[\textcolor{red}{#1}]}

\begin{document} 

\title{\boldmath The Eikonal Phase and Spinning Observables}

%% %simple case: 2 authors, same institution
%% \author{A. Uthor}
%% \author{and A. Nother Author}
%% \affiliation{Institution,\\Address, Country}

% more complex case: 4 authors, 3 institutions, 2 footnotes
%\author{Zvi Bern}
\author{Juan Pablo Gatica}

% The "\note" macro will give a warning: "Ignoring empty anchor..."
% you can safely ignore it.

\affiliation{Mani L Bhaumik Institute for Theoretical Physics,\\UCLA Department of Physics and Astronomy, Los Angeles, CA 90095 USA}

% e-mail addresses: one for each author, in the same order as the authors
%\emailAdd{bern@physics.ucla.edu, jpgatica3541@g.ucla.edu}
\emailAdd{jpgatica3541@g.ucla.edu}

\abstract{We set up the Kosower, Maybee, O'Connell formalism  for generic spin following a previous field theory construction  and use it to extract a next-to-leading order eikonal formula for the linear-in-spin momentum impulse, $\Delta p^{\mu}$, and spin kick, $\Delta S^{\mu \nu}$. We explicitly test the formalism for the case of electrodynamics and find agreement with previous QFT and worldline results. This includes the case where multiple spins propagate with allowed spin transitions, or equivalently where no spin supplementary condition is imposed on the worldline equations of motion.
}

\iffalse
We derive an eikonal formula for the tree level and one loop, that follow from the Kosower, Maybee, O'Connell (KMOC) formalism.  To verify the results of the eikonal formula, we calculate $\Delta S^{\mu \nu}$ and $\Delta p^{\mu}$ for the scattering of a massive general spin particle with a massive scalar particle interacting via QED up to $\mathcal{O}(\alpha^2 S)$ and compare to the existing literature.

We perform these calculations without the assumption of an SSC in order to explore the details of extra degrees of freedom found in general spin QFT and worldlines where.
\fi

\maketitle
\flushbottom

\newpage

\section{Introduction}
The direct observation of gravitational waves from binary mergers~\cite{LIGOScientific:2016aoc, lIGOScientific:2017vwq} has ushered in a new era of observational astronomy. As
gravitational-wave detectors become more sensitive ~\cite{Punturo:2010zz, LISA:2017pwj, Reitze:2019iox}, spin, tidal effects and absorption will become ever more important. Here we will focus on the problem of spin in the context of the Kosower, Maybee and O'Connell (KMOC) formalism~\cite{Kosower:2018adc}, which directly computes observables such as the impulse and the spin kick. We use the KMOC formalism to obtain eikonal formulae at next-to-leading order in the coupling and linear-in-spin.  We confirm that for the case of electrodynamics the derived scattering observable impulses match previously obtained results~\cite{Bern:2023ity} with or without a spin supplementary condition imposed on the worldline equations of motion or equivalently with or without spin transitions allowed in field theory.

The study of the dynamics of spinning objects in general relativity~\cite{Mathisson:1937zz, Papapetrou:1951pa, Pirani:1956tn, Tulczyjew:1959} has a long history, in both the post-Newtonian
(PN) framework~\cite{Barker:1970zr, Barker:1975ae, Kidder:1992fr, Kidder:1995zr, Blanchet:1998vx, Tagoshi:2000zg,
Porto:2005ac, Faye:2006gx, Blanchet:2006gy, Damour:2007nc, Steinhoff:2007mb, Levi:2008nh,
Steinhoff:2008zr, Steinhoff:2008ji, Marsat:2012fn, Hergt:2010pa,
 Porto:2010tr,  Levi:2010zu,  Porto:2010zg, Levi:2011eq,
 Porto:2012as,  Hergt:2012zx,  Bohe:2012mr,  Hartung:2013dza,  Marsat:2013wwa,  Levi:2014gsa,
Vaidya:2014kza, Bohe:2015ana,  Bini:2017pee,  Siemonsen:2017yux,
 Porto:2006bt,  Porto:2007tt,  Porto:2008tb,  Porto:2008jj,  Levi:2014sba,
Levi:2015msa, % this is Levi-Steinhoff paper
Levi:2015uxa,  Levi:2015ixa,  Levi:2016ofk,   Levi:2019kgk,  Levi:2020lfn,
 Levi:2020kvb,  Levi:2020uwu,  Kim:2021rfj,  Maia:2017gxn,  Maia:2017yok,  Cho:2021mqw,  Cho:2022syn,
 Kim:2022pou,  Mandal:2022nty, Kim:2022bwv, Mandal:2022ufb, Levi:2022dqm,  Levi:2022rrq},
where simultaneous expansions in Newton's constant $G$ and in the velocity $v$ are carried out, and the post-Minkowskian (PM) framework~\cite{Bini:2017xzy, Bini:2018ywr, 
Maybee:2019jus,   Guevara:2019fsj,
  Chung:2020rrz,   Guevara:2017csg,   Vines:2018gqi,  Damgaard:2019lfh,
% checked up to here
Aoude:2020onz,
Vines:2017hyw,
Guevara:2018wpp,   Chung:2018kqs,
Chung:2019duq,
Bern:2020buy,
Kosmopoulos:2021zoq,
Liu:2021zxr,
Aoude:2021oqj,
Jakobsen:2021lvp,
Jakobsen:2021zvh,
Chen:2021kxt,
Chen:2022clh,
Cristofoli:2021jas,
Chiodaroli:2021eug,
Cangemi:2022abk,
Cangemi:2022bew,
Haddad:2021znf,
Aoude:2022trd,
Menezes:2022tcs,
Bern:2022kto,
Alessio:2022kwv,
Alessio:2023kgf,
Bjerrum-Bohr:2023jau,
Damgaard:2022jem,
Haddad:2023ylx,
Aoude:2023vdk,
Jakobsen:2023ndj,
Jakobsen:2023hig,
Heissenberg:2023uvo,
Bianchi:2023lrg},
where the complete velocity dependence is kept while an expansion is carried out in Newton's constant.  The electromagnetic case is similar~\cite{Westpfahl:1979gu, Damour:1990jh, Buonanno:2000qq, Kosower:2018adc, Saketh:2021sri, Bern:2021xze, Bern:2023ccb}, with the post-Coulombian (PC) and post-Lorentzian (PL) expansions corresponding to the gravitational PN and PM expansions. Calculation of observables in these expansions have been done in numerous formalisms such as the eikonal formalism \cite{DiVecchia:2021bdo, DiVecchia:2022piu, DiVecchia:2023frv}, which has the promise of directly linking observables to the eikonal phase in a simple way.

Including spin in the aforementioned calculations has been a rich topic in the PM and PL expansions; calculations have been performed for the momentum impulse and the spin kick using fixed spin states \cite{Maybee:2019jus, Aoude:2021oqj, FebresCordero:2022jts, Menezes:2022tcs}, and worldline QFT \cite{Jakobsen:2021zvh, Jakobsen:2022fcj, Jakobsen:2022zsx, Jakobsen:2023ndj, Jakobsen:2023hig}. More recently, work has been done in including spin to waveform, memory effect, and radiation calculations \cite{Jakobsen:2021lvp, Bini:2023mdz, Aoude:2023dui, Brandhuber:2023hhl, DeAngelis:2023lvf, Heissenberg:2023uvo}. For including spin in the momentum impulse and the spin kick, much progress has been made in performing these calculations using general spin QFT \cite{Bern:2020buy, Kosmopoulos:2021zoq, Bern:2023ity}. Traditionally, massive higher spin \cite{Singh:1974qz, Lindwasser:2023dcv, Lindwasser:2023zwo, Cangemi:2022abk, arkanihamed2021scattering, Johansson:2019dnu} requires the inclusion of lower spin auxiliary fields to eliminate nonphysical degrees of freedom; however, these introduce complicated propagator structures. The general spin QFT avoids the complicated propagator structure by using unconstrained, non-transverse fields that are spin-s representations of the Lorentz group; however, multiple spins can propagate with allowed transitions. 

An interesting puzzle that arises in general spin QFT when multiple spins can propagate with allowed transitions is that extra independent Wilson coefficients appear in the derived physical results~\cite{Bern:2022kto}.  Ref.~\cite{Bern:2023ity} showed that at least in the context of electrodynamics similar extra Wilson coefficients are obtained using a worldline formalism without a spin supplementary condition imposed.  Here we confirm these results using the KMOC formalism. 

The study of spinning amplitudes and their observables has also provided a greater understanding to some underlying structures in two-body scattering. In Refs.~\cite{Bern:2020buy, Kosmopoulos:2021zoq, Bern:2023ity, Cristofoli:2021jas}, it has been shown that there is a direct link between spinning scattering observables and the eikonal phase. While in Refs.~\cite{Bern:2020buy, Kosmopoulos:2021zoq} such a formula was found for the momentum impulse and spin vector impulse via ansatz, Ref.~\cite{Cristofoli:2021jas} has found a way to partially derive these formulae by directly working with the eikonal phase in the KMOC formalism. This has since left a mystery about how the complete formula in Refs.~\cite{Bern:2020buy, Kosmopoulos:2021zoq}, especially the $D_{L}(f, g)$ terms, could be derived.  

In this paper, we will calculate the momentum impulse $\Delta p^{\mu}$ and the spin tensor impulse $\Delta S^{\mu \nu}$ of a spinning body scattering off a non-spinning body to one loop, linear-in-spin order.  We propose a procedure using the non-transverse properties of general spin fields in KMOC, which will allow us to derive a formula relating the eikonal phase of general spin amplitudes to their classical observables, including $D_{L}(f, g)$-like terms. To verify our results we will plug in amplitudes for general spin field theory coupled to electrodynamics and compare to Ref.~\cite{Bern:2023ity}, which we match when taking in to account differences in convention. Throughout the paper we will be using the mostly minus metric signature.

\section{Review}
\subsection{The Classical Scaling with Spin}
We will be taking the classical limit of two-body scattering of particles mediated by electrodynamics separated by impact parameter $b$ with charge $e q_i$, momentum $p_i$, mass $m_i$ and spin $S_i$. In electrodynamics, the classical limit implies a hierarchy between the particle Compton wavelength, $\lambda_{c_i} = {\hbar}/{m_i c}$, and the classical charge radius, $r_{Q_i} = {e^2 q_i^2}/{4 \pi m_i}$; namely,  $\lambda_{c_i} \ll r_{Q_i}$, which corresponds to the large charge limit in QED \cite{Bern:2021xze, Kosower:2018adc}. We will also be performing the Post-Lorentzian (PL) expansion, whose expansion parameter $r_{Q_i}/b \ll 1$, amounts to a large separation limit. Performed together, the classical limit and PL expansion gives us the following hierarchy of length scales
\begin{equation}
    \lambda_{c_i} \ll r_{Q_i} \ll b.
\end{equation}
For scattering amplitudes, it is more natural to consider the above hierarchy in momentum space. In two-body scattering the momentum transfer $q$ is the conjugate variable to the impact parameter $b$,
\begin{equation}
    |b| \sim \frac{1}{|q|} \gg \frac{1}{m_i}, 
\end{equation}
which implies
\begin{equation}
    |q| \ll m_i \sim |p_i|,
\end{equation}
where $p$ is some external four-momentum with natural units $\hbar = c = 1$. We also know, for two-body classical systems, the angular momentum can be expressed as 
\begin{equation}
   |J_i|  \equiv |p_i| |b| \gg 1.
\end{equation}
We make the same assumption as Ref.~\cite{Bern:2020buy} for the magnitudes of the spin and angular momentum, $ |J_i| \sim |S_i|  $. The above is summarized in the following scaling, 
\begin{equation}\label{classicalscaling}
    p \rightarrow \lambda^{0} p,  \hskip .8 cm  q \rightarrow \lambda q, \hskip .8 cm  b \rightarrow \lambda^{-1} b, \hskip .8 cm  S \rightarrow \lambda^{-1} S,
\end{equation}
where $\lambda \ll 1$, which helps us organize the classical limit.

%\subsection{What do we mean by spin?}
\subsection{Definition of Spin}
We follow Refs.~\cite{Bern:2023ity, Bern:2020buy, Kosmopoulos:2021zoq}, and model the massive scattering bodies with complex general spin-s fields $\Phi^{\mu_1 \cdots \mu_s}_s = \Phi_s^{\mu(s)}$, where we choose $s \in \mathbb{Z}^{0+}$.  These fields are symmetric and traceless but not transverse; therefore we do not need to include lower-spin auxiliary fields, such as those in Refs.~\cite{Lindwasser:2023dcv, Lindwasser:2023zwo, Singh:1974qz}. It has been argued in Ref.~\cite{Bern:2023ity} that issues arising from these non-transverse states do not contribute to the classical limit. As a result, from the rank-$s$ polarization tensors associated with the non-transverse fields, we construct the following completeness relation
\begin{equation}\label{complrel}
    \delta^{\mu(s)}_{\,\,\,\,\,\nu(s)} = \sum_{a} \epsilon^{\mu(s)}_{a}(p) \epsilon^{*\, a}_{\nu(s) }(p),
\end{equation}
where we sum over the little group indices $a$.  

We use the spin tensor, which we define as
\begin{equation}\label{stensordef}
    [S^{\mu \nu}(p)]^{a'}_{\,\,a} \equiv \epsilon^{* \, a'}_{\alpha(s)}(p) \left(\mathit{M}^{\mu \nu}\right)^{\alpha(s)}_{\,\,\, \beta(s)} \epsilon^{\beta(s)}_{a}(p),
\end{equation}
where $\left(\mathit{M}^{\mu \nu}\right)^{\alpha(s)}_{\,\,\, \beta(s)}$ is the spin-$s$ representation of the Lorentz generator
\begin{equation}
    \left(\mathit{M}^{\mu \nu}\right)^{\alpha(s)}_{\,\,\, \beta(s)} = i s \, \delta^{[\mu}_{(\beta_1} \eta^{\nu] (\alpha_1} \delta^{\alpha_2}_{\beta_2} \cdots \delta^{\alpha_s)}_{\beta_s)},
\end{equation}
satisfying the Lorentz algebra
\begin{equation}
    [\mathit{M}^{\mu \nu}, \mathit{M}^{\rho \sigma}] = - i \left(\eta ^{\mu \rho} \mathit{M}^{\sigma \nu} + \eta ^{\nu \rho} \mathit{M}^{\mu \sigma} - \eta ^{\mu \sigma} \mathit{M}^{\rho \nu} - \eta ^{\nu \sigma} \mathit{M}^{\mu \rho}\right).
\end{equation}
Note that because we have constructed the spin tensor from our unconstrained polarization tensors, the spin tensor does not necessarily obey the covariant spin supplementary condition (SSC), $p_{\mu} S^{\mu \nu}(p) = 0$~\cite{Bern:2023ity}.  We relax the SSC in order to include boost degrees of freedom in the spin tensor, as in Ref.~\cite{Bern:2023ity}, although these can be set to zero at any point.

\subsection{Basics of Observable Impulses}
We now review the methods covered in Refs.~\cite{Kosower:2018adc, Herrmann:2021tct} to arrive at different observable impulses. First, we measure the expectation value of some quantum operator $\mathbb{O}$ from the asymptotic past (``in'') to the asymptotic future (``out'') and take the difference; this results in an impulse of some corresponding observable $\mathcal{O}$,
\begin{equation}\label{kmocbase}
    \Delta \mathcal{O} = \langle{\rm out}| \mathbb{O} |{\rm out} \rangle - \langle{\rm in}| \mathbb{O} |{\rm in} \rangle.
\end{equation}
We then relate these out-states to in-states by the S matrix, $S = 1 + i T$. We model our in-state by some $N$-particle wavepacket $| \Psi \rangle$, which depends on scalar wavefunctions $\phi (p_i)$, little group dependent parts of the wavefunction $\xi_{a_i}$, and momentum eigenstates $| p_i ; a_i \rangle $, where $a_i$ are little group indices. Doing so gives a natural separation that has come to be known as the ``virtual'' and ``real'' part of the impulse \cite{Kosower:2018adc, Herrmann:2021tct}
\begin{equation}\label{vandr}
    \Delta \mathcal{O} = \langle{\Psi}|  i [\mathbb{O}, T] |{\Psi} \rangle + \langle{\Psi}|  T^{\dagger} [\mathbb{O}, T] |{\Psi} \rangle,
\end{equation}
where
\begin{equation}
    | \Psi \rangle =\bigotimes_{i = 1}^{N} \sum_{a_i} \frac{1}{N!}\int d \Phi(p_i) \, \phi(p_i) \, \xi^{a_i} \, e^{i b_i \cdot p_i} \, | p_i ; a_i  \rangle.
\end{equation}
Here we write the integration over Lorentz invariant phase space as 
\begin{equation}
    \int d\Phi(p_i) \equiv \int \hat{d}^{D}p_i \, \hat{\delta}^{(+)}(p_i^2 - m_i^2),
\end{equation}
where $D = 4 - 2 \epsilon$, $\hat{d}^{D}p_i \equiv {d^{D}p_i}/{(2 \pi)^4}$, and $\hat{\delta}^{(+)}(k^2 - m^2) \equiv 2 \pi \delta^{(+)}(k^2 - m^2)$, which only takes into account positive energy states; in the future, we will drop the $(+)$ superscript on the $\delta$-function, since in the classical limit we do not have to make this distinction. We also assume a displacement $b_i$ for each single-particle wavepacket with respect to some arbitrary origin. We now fix ourselves to $N =2$ body scattering, where we take 
\begin{equation}
    \frac{|p_1;a_1\rangle \otimes |p_2;a_2 \rangle}{2!} = |p_1 p_2;a_1 a_2\rangle.
\end{equation}
Now we can write the observable impulse for two-body scattering as 
\begin{equation}\label{kmocschem}
    \Delta \mathcal{O} = \prod_{i = 1}^{2} \, \sum_{a_i, a_i'}\int d \Phi(p_i) \, d \Phi(p_i') \,\phi^{*}(p_i')\phi(p_i) \, \xi^{*}_{a_i'} \, \xi^{a_i} \, e^{-i b_i \cdot( p_i'- p_i)} \, (\mathcal{I}^{\{a'_i\}}_{v \,\, \{a_i\}}  + \mathcal{I}^{\{a'_i\}}_{r \,\, \{a_i\}}),
\end{equation}
where
\begin{equation}
    \mathcal{I}^{\{a'_i\}}_{v \,\, \{a_i\}} = \langle{p_1' p_2';a_1' a_2'}|  i [\mathbb{O}, T] |{ p_1 p_2; a_1 a_2} \rangle, \hspace{0.5cm} \mathcal{I}^{\{a'_i\}}_{r \,\, \{a_i\}} = \langle{p_1' p_2';a_1' a_2'}|  T^{\dagger} [\mathbb{O}, T] |{p_1 p_2; a_1 a_2} \rangle,
\end{equation}
are known as the virtual kernel and real kernel, respectively. 

For simplicity, we consider only one of our scattering bodies to be spinning; therefore, we can neglect one of these bodies' little group information. Using these states, we define our two-body scattering amplitude as
\begin{align}\label{ampdef}
    \langle{p_1' p_2';a_1'}|T|p_1 p_2; a_1\rangle &= \hat{\delta}^{(4)}(p_1' + p_2' - p_1 - p_2) \left[\mathcal{A}(p_1, p_2, p_1', p_2', S_1) \right]^{a_1'}_{\,\,a_1} \nonumber\\
    & \equiv \hat{\delta}^{(4)}(p_1' + p_2' - p_1 - p_2) \, \epsilon^{* \, a_1'}(p_1') \cdot \mathbb{A}(p_1, p_2, p_1', p_2') \cdot \epsilon_{a_1}(p_1),
\end{align}
where in the last line we stripped the amplitude of its polarizations, allowing us to work with the Lorentz generator as opposed to the spin tensor; we also define $S^{\mu \nu}(p_1) \equiv S^{\mu \nu}_1$. We leave the summation over representation indices in the second line of Eq.~(\ref{ampdef}) implicit.

We expand the amplitudes in powers of the coupling constant $g$ and powers of spin,
\begin{align}\label{ampexp}
    \mathcal{A}(p_i, p_i', S_1) =& g^2\left( \mathcal{A}^{(1, 0)}(p_i, p_i')+ \mathcal{A}^{(1, 1)}(p_i, p_i', S_1)+ \cdots \right) \nonumber\\
    & \null+ g^4 \left( \mathcal{A}^{(2, 0)}(p_i, p_i')+  \mathcal{A}^{(2, 1)}(p_i, p_i', S_1) + \cdots \right) + \cdots
\end{align}
where, for $\mathcal{A}^{(i, j)}$, $i$ tracks the powers of $g^2$ and $j$ tracks the powers of the spin tensor. In the case of two-body scattering of general spin fields coupled to electrodynamics, we expand in $g^2 = 4 \pi \alpha q_1 q_2$, where $\alpha = e^2/4 \pi$ is the QED fine structure constant. 
Note that our amplitude in the first line of Eq.~(\ref{ampdef}) is a matrix with respect to the little group indices, which it inherits from its dependence on $S^{\mu \nu}$. When contracted with the $\xi^{a_i}, \xi^{*}_{a_i'}$ and integrated over the initial momentum phase space, we arrive at the expectation value of the spin tensor
\begin{equation}
    \int d \Phi (p_1) \sum_{a_1 a_1'}  \xi^{*}_{a_1'} \, \xi^{a_1} \left[S^{\mu \nu}_{1} \right]^{a_1'}_{\,\,a_1} = \langle{S^{\mu \nu}_{1} } \rangle. 
\end{equation}
We are only concerned with the classical limit of these impulses. Therefore, following the arguments of Refs.~\cite{Kosower:2018adc, Cristofoli:2021jas, Aoude:2021oqj}, these wavepackets sharply peak about their classical values, amounting to the appearance of on-shell $\delta$-functions. Similarly, since we are taking the classical limit of quantum expectation values, we must take our states to be spin coherent states, i.e. they minimize the standard deviations of the spin degrees of freedom of the observables \cite{Bern:2020buy, Aoude:2021oqj}. As a result, we are allowed to simplify Eq.~(\ref{kmocschem})
\begin{equation}
    \Delta \mathcal{O} = \int \hat{d}^{D}q \, \hat{\delta}(2 p_1 \cdot q) \, \hat{\delta}(2 p_2 \cdot q) \, e^{-i b \cdot q} \,\langle{(\mathcal{I}_v + \mathcal{I}_r)}\rangle \equiv \int \cancel{D}q \, e^{-i b \cdot q} \, (\mathcal{I}_v + \mathcal{I}_r),
\end{equation}
where $q^{\mu} = p'^{\mu}_1 - p^{\mu}_1$ is the small momentum transfer conjugate to the impact parameter $b^{\mu} = b^{\mu}_2 - b^{\mu}_1$, and we have absorbed the $\delta$-functions into the the measure via the $\cancel{D}$ notation. We will leave the expectation value brackets implicit. It should be emphasized that the amplitudes, $\mathcal{A} $ or $\mathbb{A}$, spin tensor $S^{\mu \nu}$, and Lorentz generator $\mathit{M}^{\mu \nu}$ are matrix valued in their respective representation indices; in the case of $\mathcal{A}$ and $S^{\mu \nu}$, these are the little group indices $a_i, a_i'$\hspace{0.01cm}; for $\mathbb{A}$ and $\mathit{M}^{\mu \nu}$, these are the Lorentz group representation indices $\alpha(s)$. 

We will be calculating both the momentum impulse $\Delta p^{\mu}$ and the spin kick $\Delta S^{\mu \nu}$ to linear order in spin and next to leading order in the coupling constant; therefore, we need to consider the corresponding operators $\mathbb{P}^{\mu}$ and $\mathbb{S}^{\mu \nu}$, respectively, and their expectation values.

\subsection{Momentum Impulse Set Up}
Starting with the easier case, we consider the momentum impulse. The following treatment was presented in Refs.~\cite{Herrmann:2021tct, Kosower:2018adc, Maybee:2019jus, Cristofoli:2021jas, Aoude:2021oqj} for the scalar and fixed spin cases; however, we are working with general spin fields and therefore some new nuances need to be considered.

First, we consider the virtual kernel of the momentum impulse,
\begin{align}\label{Ivdp}
   \langle{\Psi}|  i [\mathbb{P}^{\mu}, T] |{\Psi} \rangle= \int \cancel{D}q \, e^{- i b \cdot q} \, \mathcal{I}^{\mu}_v &= \int \cancel{D}q \, e^{-i b \cdot q} \, i q^{\mu} \, \epsilon^{*}(p_1 + q)\cdot \mathbb{A}(p_1, p_2, q) \cdot \epsilon(p_1) \nonumber \\
   & = \int \cancel{D}q \, e^{-i b \cdot q} \, i q^{\mu} \, \epsilon^{*}(p_1 + q)\cdot \epsilon(p_1) \, \mathcal{A}(p_1, p_2, S_1),
\end{align}
where we insert a complete set of states, Eq.~(\ref{complrel}), on either side of the polarization stripped amplitude to arrive at the last line. The product of polarizations in the second line of Eq.~(\ref{Ivdp}) can be re-expressed as an exponential
\begin{equation}\label{polprodexp}
    \epsilon^{*}(p_1+q) \cdot \epsilon(p_1) = \text{exp}\left(i q_{\mu} \frac{ S^{\mu \nu}_1 (k_{\nu} + p_{1\,\nu}) }{p_1 \cdot k + m_1^2} \right) = \text{exp}(i q \cdot \omega),
\end{equation}
where $k$ is some arbitrary reference frame from which one boosts from to arrive at either momentum $p_1$ or $p_1 + q$; in our case, it is the rest frame of the mass $m_1$ body. A derivation can be found in Appendix \ref{poltenssec} as well as in Refs.~\cite{Bern:2023ity, Chung:2018kqs}. This naturally leads us to introduce a new conjugate variable to $q$, which we call the covariant impact parameter \cite{Bern:2020buy, Kosmopoulos:2021zoq}
\begin{equation}\label{bcovdef}
    b^{\mu}_{\text{cov}} = b^{\mu} - \omega^{\mu} = b^{\mu} - \frac{ S^{\mu \nu}_1 (k_{\nu} + p_{1\,\nu})}{p_1 \cdot k + m_1^2}.
\end{equation}
This leaves us with the all orders in spin and coupling constant result
\begin{equation}
     \langle{\Psi}|  i [\mathbb{P}^{\mu}, T] |{\Psi} \rangle = \int \cancel{D}q \, e^{-i b_{\text{cov}} \cdot q} \, i q^{\mu} \, \mathcal{A}(p_1, p_2, S_1).
\end{equation}
Now we consider the real kernel of the momentum impulse,
\begin{equation}
  \int \cancel{D}q \, e^{- i b \cdot q} \, \mathcal{I}^{\mu}_{r} = \langle{\Psi}| T^{\dagger} [\mathbb{P}^{\mu}_{1}, T] |\Psi \rangle  =  \bigotimes_i^{N} \int \frac{d \rho_i}{N!}  \langle{\Psi}| T^{\dagger} |\rho_i \rangle \langle{\rho_i}| [\mathbb{P}^{\mu}_{1}, T] |\Psi \rangle,
\end{equation}
where we inserted some complete set of states $\rho_i$. By choosing the content of our complete set of states, we automatically fix the order in the coupling constant $g$ to which we can calculate. By virtue of the external states, we must insert a minimum of one massive state per matter line so as to preserve the mass content. In principle, we can insert additional massless states but this leads to higher-order contributions in $g$, which we are not concerned with in this paper. Choosing the minimal complete set of states fixes us to the next to leading order in $g$ real kernel for the momentum impulse
\begin{align}\label{Irdp}
    \mathcal{I}^{(2)\, \mu}_{r} &= \int \cancel{D}l \, l^{\mu} \, \epsilon^{*}(p_1 + q)\cdot \mathbb{A}^{(1)}(q-l, p_1+l, p_2 -l) \cdot \mathbb{A}^{(1)}(l, p_1, p_2) \cdot \epsilon(p_1) \nonumber \\
    & = e^{i q \cdot \omega} \int \cancel{D}l \, l^{\mu} \, \mathcal{A}^{(1)}(q-l, p_1 + l, p_2 - l, S_1) \, \mathcal{A}^{(1)}(l, p_1, p_2, S_1) \nonumber \\
    & = \int \hat{d}^{D}l \, \, l^{\mu} \, \epsilon^{*}(p_1+q) \cdot \vcenter{\hbox{\scalebox{1.0}{\realpartNLONoInsertionNoPhoton}}} \cdot \epsilon(p_1),
\end{align}
where $l^{\mu}$ is the loop momentum. In the last line, we relate the real kernel to the two-particle unitarity cut of the polarization stripped one loop amplitude, which replaces the cut momentum with on-shell $\delta$-functions; in fact, the same ones that are in the definition of $\cancel{D}l$. 

From Eq.~(\ref{Irdp}), we can expand in orders of the spin tensor. Naively, one might expect for $\mathcal{I}^{(2, 1)\, \mu}_r$ one would only need linear-in-spin combinations of the amplitudes in Eq.~(\ref{Irdp}). However, one can decompose any product of Lorentz generators into combinations of symmetric and anti-symmetric products; in fact, one can always construct a completely antisymmetric product, which reduces to be linear in $\mathit{M}^{\mu \nu}$. For example, in the case of a quadratic product
\begin{equation}
    \mathit{M}^{\mu \nu} \mathit{M}^{\rho \sigma} = \frac{1}{2} \left \{\mathit{M}^{\mu \nu}, \mathit{M}^{\rho \sigma}\right\} + \frac{1}{2} \left [\mathit{M}^{\mu \nu}, \mathit{M}^{\rho \sigma}\right],
\end{equation}
where the curly braces are an anti-commutator and the last term, by virtue of the Lorentz algebra, is linear in the Lorentz generator. At the level of the spin tensor one can summarize this in the following spin counting
\begin{equation}\label{spinscaling}
    S^{\mu \nu} \sim \mathcal{O}(s), \hspace{1cm} [\hspace{0.5cm}, \hspace{0.5cm}] \sim \mathcal{O}(s^{-1}), \hspace{1cm} \{\hspace{0.5cm},\hspace{0.5cm}\} \sim \mathcal{O}(1),
\end{equation}
where $s$ is some spin counting parameter. 
Fixing ourselves to linear order in spin, we express the real kernel for the momentum impulse as 
\begin{align}\
    &\mathcal{I}^{(2, 1) \, \mu}_r = \nonumber \\
    & \hspace{0.9cm} e^{i q \cdot \omega} \int \cancel{D}l \, l^{\mu} \left( \mathcal{A}^{(1, 0)}(q - l)\mathcal{A}^{(1, 1)}(l, p_1, p_2, S_1) + \mathcal{A}^{(1, 1)}(q - l, p_1 + l, p_2 - l, S_1)\mathcal{A}^{(1, 0)}(l)\right. \nonumber\\
    & \left. \hspace{3.2cm} \null +\frac{1}{2}\left[\mathcal{A}^{(1, 1)}(q - l, p_1 + l, p_2 - l, S_1), \mathcal{A}^{(1, 1)}(l, p_1, p_2, S_1)\right]\right) \label{realkerdp}.
\end{align}

\subsection{Spin Kick Set Up}
Whereas the momentum operator was easy to work with because we are using momentum eigenstates, the spin operator expectation value is more complicated
\begin{equation}\label{exptS}
    \langle{p'; a'}| \mathbb{S}^{\mu \nu} |p ; a \rangle  = - \frac{1}{2} \, \hat{\delta}_{\Phi}(p' - p) \, \epsilon^{*\,a'}(p')\cdot\mathit{M}^{\mu \nu} \cdot \epsilon_{a}(p),  
\end{equation}
where $\hat{\delta}_{\Phi}(p' - p)$ is the $\delta$-function for the Lorentz invariant phase space integral such that $\int d \Phi(p) \, f(p) \, \hat{\delta}_{\Phi}(p' - p) = f(p')$. A derivation for Eq.~(\ref{exptS}) can be found in Appendix \ref{spinopsec}.

Plugging in Eq.~(\ref{exptS}) to the virtual kernel we get
\begin{align}
    \langle{\Psi}|  i [\mathbb{S}^{\mu \nu}_1, T] |{\Psi} \rangle = \int \cancel{D}q \, e^{- i b \cdot q} \, \mathcal{I}^{\mu \nu}_v &= \frac{i}{2}\int \cancel{D}q \, e^{- i b \cdot q} \, \epsilon^{*}(p_1 + q) \cdot \left[\mathbb{A}(q, p_1, p_2), \mathit{M}^{\mu \nu} \right] \cdot \epsilon(p_1) \nonumber\\
    & = \frac{i}{2}\int \cancel{D}q \, e^{- i b_{\text{cov}} \cdot q} \, \left[\mathcal{A}(q, p_1, p_2, S_1), S^{\mu \nu}(p_1) \right]. 
\end{align}
For the real kernel at next to leading order in the coupling constant to all orders in spin
\begin{align}
    \mathcal{I}^{(2)\,\mu \nu}_{r} &=\frac{1}{2} \int \cancel{D}l \, \epsilon^{*}(p_1 + q) \cdot \mathbb{A}^{(1)}(q-l, p_1+l, p_2-l) \cdot 
    \left[\mathbb{A}^{(1)}(l, p_1, p_2), \mathit{M}^{\mu \nu} \right]\cdot \epsilon(p_1) \nonumber\\
    &= \frac{e^{i q \cdot \omega}}{2} \int \cancel{D}l \, \mathcal{A}^{(1)}(q-l, p_1+l, p_2-l, S_1) 
    \left[\mathcal{A}^{(1)}(l, p_1, p_2, S_1), S^{\mu \nu}_1 \right],
\end{align}
where if we use the spin scaling in Eq.~(\ref{spinscaling}) and expand to linear order in spin counting $s$,
\begin{align}
    \mathcal{I}^{(2, 1)\,\mu \nu}_{r} =& \,\frac{e^{i q \cdot \omega}}{2} \int \cancel{D}l \left(\mathcal{A}^{(1, 0)}(q-l) 
    \left[\mathcal{A}^{(1, 1)}(l, p_1, p_2), S^{\mu \nu}_1 \right] \right. \nonumber\\ 
    &\left. \hspace{1.7cm} \null+ \frac{1}{2}\left[\mathcal{A}^{(1, 1)}(q-l, p_1+l, p_2-l, S_1),\left[\mathcal{A}^{(1, 1)}(l, p_1, p_2, S_1), S^{\mu \nu}_1\right]\right]\right).
\end{align}
%A more detailed calculation of the spin kick kernels can be found in the appendix.
Before we move on to deriving the eikonal formulae for the momentum impulse and the spin kick, we must first consider some properties of general spin amplitudes. 

\section{General Spin Amplitudes}
We use general spin fields whose specifics beyond their non-transversality are not important for our derivation. For the derivation of the eikonal formula, we will keep our amplitudes general and grounded on minimal assumptions. 

To verify our results, we use the following Lagrangian density from Ref.~\cite{Bern:2023ity},
\begin{equation}\label{lagdens}
    \mathcal{L} = \mathcal{L}_{\text{min}}+\mathcal{L}_{\text{EM}} + \sum_{i} \mathcal{L}^{(i)}_{\text{non-min}},
\end{equation}
where the sum is over powers of the Lorentz generator. In this paper we expand only to leading order in powers of the Lorentz generator
\begin{align}
    &\mathcal{L}_{\text{min}} = (D^{\mu} \Phi_s)^{\dagger} \, D_{\mu} \Phi_s - m^2 \, \Phi^{\dagger}_s \, \Phi_s, \label{minlagrangian}\\
    &\mathcal{L}_{\text{EM}} =  - \frac{1}{4} F^{\mu \nu} F_{\mu \nu}, \\
    &\mathcal{L}^{(1)}_{\text{non-min}} = C_1 F_{\mu \nu} \, \Phi^{\dagger}_s \, \mathbb{M}^{\mu \nu} \, \Phi_s + \frac{D_1}{m^2} \, F_{\mu \nu} \, \left( (D_{\rho} \Phi_s)^{\dagger} \, \mathbb{M}^{\rho \mu} \, D^{\nu} \Phi_s + (D^{\nu} \Phi_s)^{\dagger} \, \mathbb{M}^{\rho \mu} \, D_{\rho} \Phi_s\right), \label{nonminlagrangian}
\end{align}
where
\begin{equation}
    D^{\mu} \Phi_s = \partial^{\mu} \Phi_s - i \sqrt{4 \pi \alpha} \, q A^{\mu} \Phi_s,\, \,\,\,\,\, F_{\mu \nu} = 2 \, \partial_{[\mu} A_{\nu]},
\end{equation}
$\Phi_s$ are the general spin fields, $C_1$ and $D_1$ are free parameters that, in principle, depend on macroscopic properties of the scattering bodies.  In the usual worldline construction~\cite{Levi:2015msa, Steinhoff:2014kwa, Steinhoff:2015ksa, Vines:2016unv, citeycite}, where a spin supplementary condition is imposed, only the analog of the $C_1$ coefficients appears. $D_1$ is an example of an ``extra Wilson coefficient'' that appears when no SSC is imposed on the worldline, or when multiple spin states with allowed transitions propagate on field theory, as we allow here.
Summations over representation indices of the fields and the Lorentz generators are left implicit. We are not including lower spin auxiliary fields to $\Phi_s$, meaning we do not have transverse fields and therefore we are able to construct a simple completeness relation as in Eq.~(\ref{complrel}). 

We take the classical limit of two-body, general spin scattering amplitudes by using the scaling in Eq.~(\ref{classicalscaling}) and expanding in the $\lambda \rightarrow 0$ limit. We also demand that we are calculating the long-range contributions of these amplitudes, which amounts to the omission of any term that may cancel any massless exchange propagators in the two-body amplitude. To facilitate taking the classical limit of amplitudes, we use special kinematics \cite{Herrmann:2021tct, Bern:2021xze, Parra-Martinez:2020dzs}
\begin{equation}\label{specialkin}
    \Bar{p}_1 = p_1 + q/2 \,\,\,\,\,\,\,\,\,\,\, \Bar{p}_2 = p_2 - q/2 \,\,\,\,\,\,\,\,\,\,\, y = \frac{\Bar{p}_1 \cdot \Bar{p}_2}{\Bar{m}_1 \Bar{m}_2} = \Bar{u}_1 \cdot \Bar{u}_2 \,\,\,\,\,\,\,\,\,\,\, \Bar{m}_i^2 = \Bar{p}_i^2 = m_i^2 - q^2/4, 
\end{equation}
where $\bar{u}_i = \bar{p}_i/\bar{m}_i$; this simplifies the scale of our calculation by making $\bar{u}_i \cdot q = 0$. 

\subsection{Tree Level}
At tree level, we expect the the classical limit of the scalar amplitude to be of the form
\begin{align}\label{treeS0} 
    \mathbb{A}^{(1, 0)}(q) = \frac{ \mathcal{N}^{(1, 0)}}{q^2} \sim \mathcal{O}(\lambda^{-2}) ,
\end{align}
where, for Eq.~(\ref{lagdens}), $\mathcal{N}^{(1, 0)} = 4 y \Bar{m}_1 \Bar{m}_2 $. Note that the numerator is taken to be a constant and therefore the amplitude only has a dependence on the momentum transfer $q$.

For the linear-in-spin tree amplitude, we have a more complicated structure. We expect the same propagator structure as in the scalar case, however we can have the following monomials in the numerator
\begin{equation}\label{treemonomials}
     \Bar{u}_{1\,\mu} \mathit{M}^{\mu \nu} q_{\nu}, \hspace{1cm}   \Bar{u}_{2\,\mu} \mathit{M}^{\mu \nu} q_{\nu}, \hspace{1cm}  \Bar{u}_{1\,\mu} \mathit{M}^{\mu \nu} \Bar{u}_{2\,\nu}.
\end{equation}
If we expect all tree level amplitudes to scale classically i.e. $\mathcal{O}(\lambda^{-2})$ and if we require only long-range contributions, then only the first two monomials contribute to our amplitudes. With this in mind, we write our amplitude as 
\begin{align}\label{treeS1}
    \mathbb{A}^{(1, 1)}(q, \Bar{u}_1, \Bar{u}_2)  = \frac{ i q_{\nu} \mathit{M}^{\mu \nu}  \left(c_1 \Bar{u}_{1 \, \mu} + c_2 \Bar{u}_{2 \, \mu}\right)}{q^2} \equiv \frac{i q_{\nu} \mathit{M}^{\mu \nu} \mathcal{N}^{(1, 1)}_{\mu}(\Bar{u}_1, \Bar{u_2})}{q^2} , 
\end{align}
where we have assumed that the Lorentz generator scales the same way as the spin tensor since they are simply related by Eq.~(\ref{stensordef}).

\subsection{One Loop}
As has been shown in Refs.~\cite{Bern:2023ity, Forde:2007mi, Bern:2021xze}, when calculating the classical limit of the two-body, one loop amplitude using the soft expansion, it is possible to separate contributions into box diagram and triangle diagram contributions
\begin{equation}\label{integraldecomp}
    \mathcal{A}^{(2)} = c_{\Delta} I_{\Delta} + c_{\nabla} I_{\nabla} + d_{\text{B}} (I_{\text{B}} + I_{\text{xB}}) + \cdots. 
\end{equation}
The associated integrals are the following
\begin{align}
    &I_{\text{B}} = \int \hat{d}^{D} l \frac{1}{l^2 (q - l)^2 (2 \Bar{u}_1 \cdot l + i0)(-2 \Bar{u}_2 \cdot l + i0)} \sim \mathcal{O}(\lambda^{-2}), \\
    &I_{\text{xB}} = \int \hat{d}^{D} l \frac{1}{l^2 (q - l)^2 (2 \Bar{u}_1 \cdot l + i0)(2 \Bar{u}_2 \cdot l + i0)} \sim \mathcal{O}(\lambda^{-2}) , \\
    &I_{\Delta} = \int \hat{d}^{D} l \frac{1}{l^2 (q - l)^2 (2 \Bar{u}_1 \cdot l + i0)} \sim \mathcal{O}(\lambda^{-1}), \\
    &I_{\nabla} = \int \hat{d}^{D} l \frac{1}{l^2 (q - l)^2 (-2 \Bar{u}_2 \cdot l + i0)} \sim \mathcal{O}(\lambda^{-1}), 
\end{align}
where we see two different scalings for the same amplitude. The triangle contributions have the expected one loop classical scaling, $\mathcal{O}(\lambda^{-1})$. On the other hand, the box integral terms scale lower than the expected classical scaling; this is known as super-classical or classically singular scaling \cite{Kosower:2018adc, Herrmann:2021tct, Bern:2021xze}. 

The classically singular parts of the one loop amplitude, $\mathcal{O}(\lambda^{-2})$, carry an infrared divergence that cancels non-trivially when calculating observables. Therefore, it is convenient to systematically separate classically singular and classical terms in the one loop amplitude. We do this by separating the one loop amplitude into real and imaginary parts 
\begin{equation}
    \mathcal{A}^{(2)} = \text{Re} \, \mathcal{A}^{(2)} + i \, \text{Im} \, \mathcal{A}^{(2)},
\end{equation}
and using unitarity,
\begin{equation}\label{opticaltheorem}
   2 \text{Im} \left[\vcenter{\hbox{\scalebox{1.0}{ \kmocvirtualnlonolab}}}\right] =  \int d \Phi_2 \vcenter{\hbox{\scalebox{1.0}{\realpartNLONoInsertionNoPhoton}}},
\end{equation}
where the integration is over the complete set of states inserted between the two tree level amplitudes. It can be shown that the combination of classically singular terms in Eq.~(\ref{integraldecomp}) is contained in the imaginary part of the one loop amplitude, therefore the real part is the combination of classical terms i.e. the triangle contributions. In the case of scalar particles, the imaginary part is exactly the classically singular term; the situation is less straightforward when we introduce spin.

We now consider $\text{Im}\,\mathbb{A}^{(2, 1)}$ explicitly
\begin{align}\label{optthmbase}
    2 \,\text{Im} \, &\mathbb{A}^{(2, 1)}(q, p_1, p_2) \nonumber\\
     &=   \int \cancel{D} l \left(\mathbb{A}^{(1, 1)}(q - l, p_1 + l, p_1 - l)\mathbb{A}^{(1, 0)}(l) + \mathbb{A}^{(1, 0)}(q - l)\mathbb{A}^{(1, 1)}(l, p_1, p_2) \right. \nonumber\\
    & \left. \hspace{1.7cm} \null + \frac{1}{2} \left[ \mathbb{A}^{(1, 1)}(q - l, p_1 + l, p_2 - l), \mathbb{A}^{(1, 1)}(l, p_1, p_2)\right] \right).
\end{align}
We are interested in the amplitude with special kinematic variables, Eq.~(\ref{specialkin}); therefore, we have to consider shifts in the external momentum by the small momentum transfer $q$ as well as the shifts coming from the amplitudes with propagator $(q - l)^{-2}$. These shifts cause the promotion of certain terms from classically singular to classical contributions. It should be noted that the last line in Eq.~(\ref{optthmbase}) already scales classically and therefore is not promoted. The promotions come from the following transformations
\begin{align}
    &\mathbb{A}^{(1, 1)}(l, p_1, p_2) \rightarrow \left(1 - \left(\frac{q - l}{2}\right) \cdot \left(\frac{\partial}{\partial \Bar{p}_1} - \frac{\partial}{\partial \Bar{p}_2}\right)\right)\mathbb{A}^{(1, 1)}(l, \Bar{p}_1, \Bar{p}_2), \label{leftderiv}\\
    &\mathbb{A}^{(1, 1)}(q - l, p_1 + l, p_2 - l) \rightarrow \left(1 + \frac{l}{2} \cdot \left(\frac{\partial}{\partial \Bar{p}_1} - \frac{\partial}{\partial \Bar{p}_2}\right)\right)\mathbb{A}^{(1, 1)}(q-l, \Bar{p}_1, \Bar{p}_2). \label{rightderiv}
\end{align}
We also exploit a feature of the combination of amplitudes we are working with,
\begin{align}
    &\mathbb{A}^{(1, 0)}(q - l) \left(\frac{q - l}{2}\right) \cdot \left(\frac{\partial}{\partial \Bar{p}_1} - \frac{\partial}{\partial \Bar{p}_2}\right)\mathbb{A}^{(1, 1)}(l, \Bar{p}_1, \Bar{p}_2) = \nonumber \\
    & \null - \mathbb{A}^{(1, 0)}(l)\frac{l}{2} \cdot \left(\frac{\partial}{\partial \Bar{p}_1} - \frac{\partial}{\partial \Bar{p}_2}\right)\mathbb{A}^{(1, 1)}(q - l, \Bar{p}_1, \Bar{p}_2),
\end{align}
to rewrite Eq.~(\ref{optthmbase}) in terms of the special kinematics variables,
\begin{align}
    2 \, \text{Im}\, &\mathbb{A}^{(2, 1)}(q, p_1, p_2)
     =   \int \cancel{D} l \left(\mathbb{A}^{(1, 1)}(q - l, \Bar{p}_1, \Bar{p}_2)\mathbb{A}^{(1, 0)}(l) + \mathbb{A}^{(1, 0)}(q - l)\mathbb{A}^{(1, 1)}(l, \Bar{p}_1, \Bar{p}_2) \right. \nonumber\\
    & \left. \hspace{4.3cm} \null + \frac{1}{2} \left[ \mathbb{A}^{(1, 1)}(q - l, \Bar{p}_1, \Bar{p}_2), \mathbb{A}^{(1, 1)}(l, \Bar{p}_1, \Bar{p}_2)\right] \right. \nonumber\\
    & \left. \hspace{4.3cm} \null + l \cdot \left(\frac{\partial}{\partial \Bar{p}_1} - \frac{\partial}{\partial \Bar{p}_2}\right) \mathbb{A}^{(1, 1)}(q-l, \Bar{p}_1, \Bar{p}_2) \mathbb{A}^{(1, 0)}(l)\right). \label{unitarityS1} 
\end{align}
Now only the top line of Eq.~(\ref{unitarityS1}) is classically singular while the rest of the terms scale classically; later on, these same shifts and transformations will occur in the real kernels of the momentum impulse and the spin kick.

\section{The Eikonal Formula}
Before we begin deriving the eikonal formula, we first briefly review the eikonal phase. A detailed discussion on the eikonal phase can be found in Refs.~\cite{DiVecchia:2021bdo, DiVecchia:2022piu, DiVecchia:2023frv}. In impact parameter space, one can write the following exponentiation of the two-body scattering amplitude
\begin{equation}\label{eikexp}
    1 + i \, \mathcal{A}(b) = \left(1 +  i \Delta(b)\right)e^{i \delta(b)},
\end{equation}
where $\delta(b)$ is the eikonal phase, $\Delta(b)$ is a quantum remainder, and
\begin{equation}
    \mathcal{A}(b) = \int \cancel{D}q \, \mathcal{A}(q) \, e^{- i b \cdot q} \equiv \text{FT}[\mathcal{A}(q)],
\end{equation}
where in a center of mass momentum configuration, with total energy $E$ and center of mass momentum $\Vec{p}$,  
\begin{equation}
    \delta(2 \Bar{p}_1 \cdot q) \delta(2 \Bar{p}_2 \cdot q) \rightarrow \frac{1}{4 |\Vec{p}| E} \delta(q^{0})\delta(q^{i}) = \frac{1}{4 \bar{m}_1 \bar{m}_2 \sqrt{y^2 - 1}} \delta(q^{0})\delta(q^{i}),
\end{equation}
where $i = 1, 2, 3$. We are suppressing the amplitude's dependence on other parameters such as the momentum or masses, for the sake of clarity. By expanding Eq.~(\ref{eikexp}) in powers of the coupling constant $g$ and in powers of classical scaling parameter $\lambda$, one can relate the classical scaling amplitude order by order to the eikonal phase. To next to leading order in the coupling constant,
\begin{equation}\label{eiktoamps}
    \delta^{(1)}(b) = \text{FT}[\mathcal{A}^{(1)}(q)], \hspace{1cm} \delta^{(2)}(b) = \text{FT}[\text{Re}\, \mathcal{A}^{(2)} ]. 
\end{equation}
The relationship between the eikonal phase and classical amplitudes becomes more complicated at higher orders in the coupling constant; however, this is beyond the scope of this paper. We assume that Eq.~(\ref{eiktoamps}) holds to all orders in the spin expansion. 

In Ref.~\cite{Bern:2020buy}, based on the structure of their results for the momentum impulse and spin kick to one loop order, the authors introduced an ansatz relating the eikonal phase to the observables in the center of mass frame, which can be expressed as
\begin{align}
    &\Delta \mathbf{p}_1 = \frac{\partial \delta}{\partial \mathbf{b}} + \frac{1}{2} \left\{ \delta, \frac{\partial \delta}{\partial \mathbf{b}}\right\} + \mathcal{D}_{L} \left(\delta, \frac{\partial \delta}{\partial \mathbf{b}}\right)  - \frac{1}{2} \frac{\partial}{\partial \mathbf{b}} \mathcal{D}_{L}(\delta, \delta) - \frac{\mathbf{p}}{2 \mathbf{p}^2} \left(\frac{\partial \delta}{\partial \mathbf{b}}\right)^2 + \mathcal{O}(\delta^3), \label{DSLformdp}\\
    &\Delta \mathbf{S}_1 = \left\{\delta, \mathbf{S}_1\right\} + \frac{1}{2}\left\{\delta, \left\{\delta, \mathbf{S}_1\right\}\right\} + \mathcal{D}_{L}(\delta, \left\{\delta, \mathbf{S}_1\right\}) - \frac{1}{2}\left\{\mathcal{D}_{L}(\delta, \delta), \mathbf{S}_1 \right\} + \mathcal{O}(\delta^3)\label{DSLformds},
\end{align}
where the bold variables are three-vectors, $\mathbf{p}$ is the center of mass momentum, $\mathbf{b}$ is the canonical impact parameter, and $\mathbf{S}$ is the rest frame spin three-vector. The spin three-vector is defined as 
\begin{equation}
    S^{i} = \frac{1}{2} \epsilon^{i j k} S_{j k},
\end{equation}
where $i, j, k = 1, 2, 3$ and we use $\epsilon^{1 2 3}$ = 1 as our convention for the Levi-Civita tensor. $\mathcal{D}_L$ is defined as 
\begin{equation}
    \mathcal{D}_L(f, g) \equiv - \epsilon_{i j k} S^{i}_{\,1} \frac{\partial f}{\partial S_{1 \, j}} \frac{\partial g}{\partial L_{k}},
\end{equation}
where we are ignoring the explicit inclusion of boost degrees of freedom since we are not making this distinction. 
The brackets are defined by the $SO(3)$ algebra
\begin{equation}
    \left\{S^{i}, S^{j} \right\} = \epsilon^{i j k} S_k,
\end{equation}
and $L_k$ is the orbital angular momentum $L^{i} = \epsilon^{i j k} \, b_{j} \, p_{k}$.

As of the writing of this paper, Eqs.~(\ref{DSLformdp}) and (\ref{DSLformds}) have been verified to quadratic order in spin, next to leading order in the coupling constant for QED and GR \cite{Bern:2023ity, Kosmopoulos:2021zoq}; however, there is no derivation. While we will be deriving an eikonal formula, we will not be deriving Eqs.~(\ref{DSLformdp}) and (\ref{DSLformds}) exactly because we will not be fixing our system to the center of mass frame and we will not explicitly break covariance. Nonetheless, we will derive an eikonal formula similar to Eqs.~(\ref{DSLformdp}) and (\ref{DSLformds}) that are written in terms of the spin tensor as opposed to the spin vector and in a covariant form. 

\subsection{Eikonal Formula Derivation}
We now derive an eikonal formula for the momentum impulse and the spin kick. We start with the comparatively simpler, linear-in-spin momentum impulse starting at tree level and then move to the one loop correction. We then do the same for the linear-in-spin spin kick. We compare all calculations to the results in \cite{Bern:2023ity} by applying the two-body scattering amplitudes associated with Eq.~(\ref{lagdens}). We expand the general spin QED impulses in the following manner
\begin{equation}
    \Delta \mathcal{O} = \alpha q_1 q_2(\Delta^{(1, 0)}\mathcal{O} +  \Delta^{(1, 1)}\mathcal{O} + \cdots ) + (\alpha q_1 q_2)^2 (\Delta^{(2, 0)}\mathcal{O} +  \Delta^{(2, 1)}\mathcal{O} + \cdots ) + \cdots 
\end{equation}
where the superscript count in the same way as in the amplitudes in Eq.~(\ref{ampexp}).

\subsection{Momentum Impulse}
\subsubsection{Tree Level}
At tree level, the momentum impulse only receives contributions from the virtual kernel. Also, because the virtual kernel already scales classically, the momentum shift to special kinematics does not result in new promoted contributions. Therefore, we have the following result to all orders in spin
\begin{align}\label{dpcov}
    \Delta^{(1)}p^{\mu}_1 &= \int \cancel{D} q \, e^{-i  b_{\text{cov}}\cdot q} \, i q^{\mu} \, \mathcal{A}^{(1)}(q, \bar{p}_1, \bar{p}_2, S_1) \nonumber \\
    &=-\frac{\partial}{\partial b_{\mu}}\int \cancel{D} q e^{-i  b_{\text{cov}}\cdot q} \, \mathcal{A}^{(1)}(q, \bar{p}_1, \bar{p}_2, S_1),
\end{align}
where the singular superscript counts the powers in the coupling constant. We can now identify the Fourier transform with respect to the new impact parameter $b_{\text{cov}}$ as the ``covariant'' eikonal phase
\begin{equation}
    \Delta^{(1)}p^{\mu}_{1} = - \frac{\partial}{\partial b_{ \mu}} \delta^{(1)}_{\text{cov}},
\end{equation}
where $\delta(b_{\text{cov}}, \bar{u}_1, \bar{u}_2, S_1) \equiv \delta_{\text{cov}} $.

Fixing to the linear-in-spin momentum impulse and using the linear-in-spin tree level amplitude for general spin QED, 
\begin{equation}\label{treelevelspinamp}
    \mathbb{A}^{(1, 1)}(q, \bar{u}_1, \bar{u}_2) = \frac{i q_{\beta} \mathit{M}^{\alpha \beta} (y D_1 \bar{u}_{1 \, \alpha} - C_1 \bar{u}_{2 \, \alpha} )}{q^2},
\end{equation}
we compute the result
\begin{equation}
    \Delta^{(1, 1)}p^{\mu}_{1} = \frac{2}{ \Bar{m}_1 \sqrt{y^2 - 1}} \frac{S_{1}^{\nu \rho}}{|b_{\text{cov}}|^2} \left(\Pi^{\mu}_{\,\,\nu} + \frac{2 \, b^{\mu}_{\text{cov}} \, b_{\text{cov}\, \nu}}{|b_{\text{cov}}|^2}\right)\left(y D_1  \Bar{u}_{1 \, \rho} - C_1 \Bar{u}_{2 \, \rho}\right),
\end{equation}
which matches the result in Ref.~\cite{Bern:2023ity}. Here $|b_{\text{cov}}|^2 = - b_{\text{cov}}^2$ and $\Pi^{\mu}_{\,\,\nu}$ is the projective derivative of the impact parameter
\begin{equation}
     \Pi^{\mu}_{\, \, \nu} = \delta^{\mu}_{\,\,\nu} + \frac{1}{y^2 - 1} \left(\bar{u}^{\mu}_1 \check{\bar{u}}_{1 \, \nu} + \bar{u}^{\mu}_2 \check{\bar{u}}_{2 \, \nu}\right),
\end{equation}
where $\check{\bar{u}}_{1, 2} \equiv \bar{u}_{1, 2} - y \, \bar{u}_{2, 1}$. To satisfy the orthogonality constraints of $b^{\mu}$ with respect to the scattering plane, i.e. $\bar{u}_i \cdot b = 0$, we require the projector $\Pi^{\mu}_{\,\, \nu}$ in place of the naive derivative, which does not take this into account.  

\subsubsection{One Loop}
At one loop, we must include the real kernel in order to ensure that all classically singular terms cancel. In the same way as $\text{Im}\,\mathbb{A}^{(2, 1)}$, we need to change to special kinematic variables in Eq.~(\ref{realkerdp}) and keep track of terms promoted to classical scaling. This results in

\begin{align}\label{realkerdpbar}
    \mathcal{I}^{(2, 1) \, \mu}_r = & \int \cancel{D}l \, \, l^{\mu} \, \epsilon^{*}(p_1 + q) \cdot  \left\{\mathbb{A}^{(1, 0)}(q - l)\mathbb{A}^{(1, 1)}(l, \Bar{p}_1, \Bar{p}_2) + \mathbb{A}^{(1, 1)}(q - l, \Bar{p}_1, \Bar{p}_2)\mathbb{A}^{(1, 0)}(l)\right. \nonumber \\
    & \left. \hspace{3.3cm} \null+ \mathbb{A}^{(1, 0)}(l) \, l \cdot \left(\frac{\partial}{\partial \Bar{p}_1} - \frac{\partial}{\partial \Bar{p}_2}\right)\mathbb{A}^{(1, 1)}(q-l, \Bar{p}_1, \Bar{p}_2)\right. \nonumber\\
    &\left. \hspace{3.3cm} \null+\frac{1}{2}\left[\mathbb{A}^{(1, 1)}(q - l, \Bar{p}_1, \Bar{p}_2), \mathbb{A}^{(1, 1)}(l, \Bar{p}_1, \Bar{p}_2)\right] \right\} \cdot \epsilon(p_1),
\end{align}
which is near identical to Eq.~(\ref{unitarityS1}) except for the explicit factor of $l^{\mu}$. In fact, when combining the virtual kernel, $\mathcal{I}^{(2, 1) \, \mu}_v$,  with Eq.~(\ref{realkerdpbar}) and focusing on the classically singular pieces, we observe the following
\begin{align}\label{tbcancelleddp}
    \int \cancel{D}q \, e^{-i q \cdot b_{\text{cov}}} \int \cancel{D} l  \left(\frac{2 l^{\mu} - q^{\mu}}{2} \right) & \left\{\mathcal{A}^{(1, 0)}(q - l)\mathcal{A}^{(1, 1)}(l, \Bar{p}_1, \Bar{p}_2, S_1) \right. \nonumber\\ 
    &\left. \hspace{0.5cm} \null+ \mathcal{A}^{(1, 1)}(q - l, \Bar{p}_1, \Bar{p}_2, S_1)\mathcal{A}^{(1, 0)}(l)\right\}. 
\end{align}
The cancellation of the classically singular pieces must occur upon integration; to make this manifest at the integrand level we make the following decomposition of factors of loop momentum vector
\begin{equation}\label{vecdecomp}
    l^{\mu} \rightarrow \frac{q^{\mu}}{2} - \frac{\left(l \cdot \bar{u}_1 \, \check{\bar{u}}^{\mu}_1 + l \cdot \bar{u}_2 \, \check{\bar{u}}^{\mu}_2\right)}{y^2 - 1} ,
\end{equation}
and apply the $\delta$-functions in the integrand, $\hat{\delta}(2 \bar{p}_{1} \cdot l - q \cdot l)\hat{\delta}(2 \bar{p}_{2} \cdot l + q \cdot l)$. Because we are only interested in long range interactions, we can make the replacement $q \cdot l \rightarrow q^2/2$; this results in the promotion of the remaining term in Eq.~(\ref{tbcancelleddp}) to scale classically. We now have a uniformly classical scaling expression for the momentum impulse
\begin{align}
    \Delta^{(2, 1)}p^{\mu}_1 = & \int \cancel{D} q \, e^{- i q \cdot b_{\text{cov}}} \, i q^{\mu} \, \text{Re} \, \mathcal{A}^{(2, 1)}(q, \Bar{u}_1, \Bar{u}_2, S_1)  \nonumber\\
     \null+  &\int \cancel{D} q e^{- i q \cdot b_{\text{cov}}} \int \cancel{D}l \left\{\left(\frac{2l^{\mu} - q^{\mu}}{4}\right)\left[\mathcal{A}^{(1, 1)}(q - l, \Bar{u}_1, \Bar{u}_2,  S_1), \mathcal{A}^{(1, 1)}(l, \Bar{u}_1, \Bar{u}_2,  S_1)\right]  \right.  \nonumber\\
    & \left. \hspace{1.2cm} \null+\mathcal{A}^{(1, 0)}(l) \left(\frac{2l^{\mu} - q^{\mu}}{2}\right) l^{\alpha} \left(\frac{1}{\bar{m}_1}\frac{\partial}{\partial \Bar{u}_{1}^{\alpha}} - \frac{1}{\bar{m}_2}\frac{\partial}{\partial \Bar{u}_{2}^{\alpha}}\right) \mathcal{A}^{(1, 1)}(q-l, \Bar{u}_1, \Bar{u}_2,  S_1) \right.  \nonumber\\
    &  \left. \hspace{1.2cm} \null+ \frac{-q^2}{4 (y^2 - 1)} \left(\frac{\check{\bar{u}}^{\mu}_1}{\bar{m}_1} - \frac{\check{\bar{u}}^{\mu}_2}{\bar{m}_2}\right) \left(\mathcal{A}^{(1, 0)}(q - l)\mathcal{A}^{(1, 1)}(l, \Bar{u}_1, \Bar{u}_2,  S_1) \right. \right. \nonumber\\
    &\left.\left. \hspace{5.8cm} \null+ \mathcal{A}^{(1, 1)}(q - l, \Bar{u}_1, \Bar{u}_2,  S_1)\mathcal{A}^{(1, 0)}(l)\right) \right\}. \label{dpLong}
\end{align}
To be able to re-express the amplitudes that depend on the loop momentum in terms of their respective eikonal phases, we make the shift $q \rightarrow q + l$ and substitute explicit factors of $q, l \rightarrow i \frac{\partial}{\partial b}$,
\begin{align}\label{dpeik}
    \Delta^{(2, 1)} p^{\mu}_{1} = & - \frac{\partial \, \delta_{\text{cov}}^{(2, 1)}}{\partial b_{ \mu}}  +  \frac{i}{2} \left[\delta_{\text{cov}}^{(1, 1)}, \frac{\partial \delta_{\text{cov}}^{(1, 1)} }{\partial b_{ \mu}}\right] + \frac{1}{(y^2 - 1)} \left(\frac{\check{\bar{u}}^{\mu}_1}{\bar{m}_1} - \frac{\check{\bar{u}}^{\mu}_2}{\bar{m}_2}\right) \frac{\partial \delta_{\text{cov}}^{(1, 0)}}{\partial b} \cdot \frac{\partial \delta_{\text{cov}}^{(1, 1)}}{\partial b} \nonumber \\
    & \null+ \frac{1}{2}\left[ \frac{\partial}{\partial b_{\mu}},  \left(\frac{1}{\Bar{m}_1}\frac{\partial}{\partial \Bar{u}_{1}^{\alpha}} - \frac{1}{\Bar{m}_2}\frac{\partial}{\partial \Bar{u}_{2}^{\alpha}} \right) \delta_{\text{cov}}^{(1, 1)} \right]\frac{\partial \delta_{\text{cov}}^{(1, 0)}}{\partial b_{ \alpha}}.
\end{align}
To verify Eq.~(\ref{dpeik}), we plug in the appropriate tree level amplitudes and linear-in-spin, one loop amplitude,
\begin{align}
    \text{Re} \mathbb{A}^{(2, 1)} =& \mathbb{A}_{\triangle + \triangledown} (q, \Bar{u}_1, \Bar{u}_2) = \frac{1}{4 \sqrt{-q^2}} i q_{\nu} \mathbb{M}^{\mu \nu}\left( \beta_{(1)} \Bar{u}_{1\, \mu} + \beta_{(2)}\Bar{u}_{2 \, \mu}\right), \label{oneloopamp}\\
    \nonumber\\
    \beta_{(1)} =& \frac{-1}{(y^2 - 1)\Bar{m}_1} \left(\left[(y^2 + 1)C_1 + (y^2 - 1) D_1 \right] \Bar{m}_1 \right. \nonumber\\ 
    &\left. \hspace{2cm} \null+ \left[C_1^2 - (y^2 + 1) C_1 D_1 + y^2 D_1^2 + (3 y^2 - 1) D_1\right] \Bar{m}_2\right), \\
    \nonumber \\
    \beta_{(2)} =& \frac{y}{(y^2 - 1)\Bar{m}_1} \left(2 C_1 \Bar{m}_1 + \left[C_1^2 - 2 C_1 D_1 + D_1^2 + 2 D_1\right]\Bar{m}_2\right),
\end{align}
and compute the result
\begin{align}
    \Delta^{(2, 1)}p^{\mu}_1 =& 
%%%%% begin : DeltaP[2,1]
    \, \frac{\pi {} \left(3 \, b^{\mu}_{\text{cov}} \, b_{\text{cov} \, \nu} + |b_{\text{cov}}|^2 \, \Pi^{\mu}_{\,\,\nu} \right)S^{\nu \rho}_1 }{2 |b_{\text{cov}}|^5 (y^2 - 1)^{3/2} \, \Bar{m}_1^2 \, \Bar{m}_2 } \left\{ \check{\Bar{u}}_{1 \, \rho}[\Bar{m}_1 (D_1 - C_1) - \Bar{m}_2 (C_1^2 - C_1 D_1 - D_1)]  \right. \nonumber \\ 
    & \left. \hspace{5.4cm} \null+ y  {} \check{\Bar{u}}_{2 \, \rho}[\Bar{m}_1 (C_1 + D_1) - \Bar{m}_2 D_1 (C_1 - D_1 - 3)] \right\} \nonumber \\
    & \null-  \frac{2 b^{\nu}_{\text{cov}} S_{1\, \nu \rho}\Pi^{\mu}_{\,\, \rho}}{ |b_{\text{cov}}|^4(y^2 - 1) \, \Bar{m}_1^2 \, \Bar{m}_2}  \left\{ 2 C_1 y \Bar{m}_1 + \Bar{m}_2 (C_1^2 - y^2 D_1(2 C_1 - D_1 - 2 )) \right\} \nonumber\\
    & \null+ \frac{4 y {} (\Bar{m}_2 \check{\Bar{u}}_1^{\mu} - \Bar{m}_1 \check{\Bar{u}}_2^{\mu}) b^{\nu}_{\text{cov}} S_{1 \, \nu \rho} }{|b_{\text{cov}}|^4 (y^2 - 1)^3 \, \Bar{m}_1^2 \, \Bar{m}_2} \left \{ \check{\Bar{u}}_1^{\rho} (D_1 - C_1) + \check{\Bar{u}}_2^{\rho}(y^2 D_1 - C_1) \right\}
%%%%% end : DeltaP[2,1]
    , 
\end{align}
where we match the result in Ref.~\cite{Bern:2023ity}, taking into account our impact parameters point in opposite directions. 

We notice a similarity between Eq.~(\ref{dpeik}) and Eq.~(\ref{DSLformdp}). The first obvious caveat is that we express our result in terms of the spin tensor as opposed to Refs.~\cite{Bern:2020buy, Bern:2023ity, Cristofoli:2021jas, Kosmopoulos:2021zoq}, which use the spin three-vector. The second is that in Eq.~(\ref{dpeik}) we work with the covariant impact parameter whereas Refs.~\cite{Bern:2020buy, Bern:2023ity, Cristofoli:2021jas, Kosmopoulos:2021zoq} use the canonical rest frame impact parameter, which are related by Eq.~(\ref{bcovdef}). Nevertheless, structural similarities are hard to not notice when comparing to Eq.~(\ref{DSLformdp}), particularly in the terms independent of $\mathcal{D}_L(f, g)$. Because we have verified Eq.~(\ref{dpeik}) using the result in Ref.~\cite{Bern:2023ity}, we conclude that we have derived an eikonal formula for the one loop, linear-in-spin momentum impulse that takes into account the contributions of the $\mathcal{D}_{L}(f, g)$ terms in Refs.~\cite{Bern:2023ity, Bern:2020buy, Kosmopoulos:2021zoq}.    

\subsection{Spin Kick}
\subsubsection{Tree Level}
Same as in the momentum impulse case, for the tree level spin kick we only need the contribution from the virtual kernel. We also do not have to worry about any promotions coming from the shift when changing to special kinematics variables because the tree level amplitude already scales classically
\begin{equation}\label{dsmntree}
    \Delta^{(1)}S^{\mu \nu}_1 = \frac{i}{2}\int \cancel{D}q \, e^{-i b_{\text{cov}} \cdot q}\left[\mathcal{A}^{(1)}(q, \bar{u}_1, \bar{u}_2, S_1), S^{\mu \nu}_1 \right].
\end{equation}
Re-expressing Eq.~(\ref{dsmntree}) in terms of the eikonal phase and covariant coordinates we get
\begin{equation}\label{dsmntreeeik}
    \Delta^{(1)}S^{\mu \nu}_{1} = \frac{i}{2}\left[\delta^{(1)}_{\text{cov}}, S^{\mu \nu}_1 \right].
\end{equation}
We check Eq.~(\ref{dsmntreeeik}) by fixing ourselves to linear order in spin and plugging in Eq.~(\ref{treelevelspinamp}) for the amplitude
\begin{equation}\label{dstreeres}
    \Delta^{(1, 1)}S^{\mu \nu}_1  = \frac{4}{ \Bar{m}_1 \sqrt{y^2 - 1}}\frac{\left(y D_1 b_{\text{cov} \, [\sigma} \Bar{u}_{1 \, \rho]} - C_1 b_{\text{cov}\, [\sigma} \Bar{u}_{2 \, \rho]}\right) \eta^{\rho [\mu} S^{\nu] \sigma}_1}{|b_{\text{cov}}|^2},
\end{equation}
where the brackets around the indices signals their anti-symmetrization, 
\begin{equation}
    a_{[\sigma} c_{\rho]} \equiv \frac{1}{2} \left(a_{\sigma} c_{\rho}  - a_{\rho} c_{\sigma}\right),
\end{equation}
for some four-vectors $a, c$, and $\eta^{\rho \mu}$ is the Minkowski metric. Eq.~(\ref{dstreeres}) matches the result in Ref.~\cite{Bern:2023ity}  taking into account our impact parameters point in opposite directions and an overall factor of 1/2, which comes from our definition of the spin operator expectation value. 

\subsubsection{One Loop}
Same as in the momentum impulse case, for the one loop calculation we need to include the real kernel contribution. After shifting to special kinematics and making explicit the classical scaling, we have the following real kernel contribution to the spin kick
\begin{align}\label{realkerndsbar}
    \mathcal{I}^{(2, 1)\, \mu \nu}_{r} =& \frac{e^{i q \cdot \omega}}{2}\int \cancel{D}l \, \left\{\mathcal{A}^{(1, 0)}(q - l) \left[ \mathcal{A}^{(1, 1)}(l, \bar{u}_1, \bar{u}_2, S_1),S_1^{\mu \nu} \right] \right. \nonumber \\
    &\left. \hspace{2cm} \null+ \mathcal{A}^{(1, 0)}(l) \frac{l^{\alpha}}{2} \left(\frac{1}{\bar{m}_1}\frac{\partial}{\partial \bar{u}^{\alpha}_1} - \frac{1}{\bar{m}_2}\frac{\partial}{\partial \bar{u}^{\alpha}_2}\right) \left[\mathcal{A}^{(1, 1)}(q-l, \bar{u}_1, \bar{u}_2, S_1),S_1^{\mu \nu}\right] \right. \nonumber\\
    & \left. \hspace{2cm} \null+ \frac{1}{2} \left[\mathcal{A}^{(1, 1)}(q-l, \bar{u}_1, \bar{u}_2, S_1),\left[\mathcal{A}^{(1, 1)}(l, \bar{u}_1, \bar{u}_2, S_1), S_1^{\mu \nu}\right]\right]\right\},
\end{align}
where only the top line of Eq.~(\ref{realkerndsbar}) has classically singular scaling. The cancellation of the classically singular terms in the real kernel and $\text{Im}\,\mathbb{A}^{(2, 1)}$ is more subtle in the spin kick case; namely, when attempting to cancel the classically singular terms, we end up with the following
\begin{equation}
    \frac{e^{i q \cdot \omega}}{4} \int \cancel{D}l\,\left[\mathcal{A}^{(1, 0)}(q - l) \mathcal{A}^{(1, 1)}(l, \bar{u}_1, \bar{u}_2, S_1) - \mathcal{A}^{(1, 0)}(l) \mathcal{A}^{(1, 1)}(q-l, \bar{u}_1, \bar{u}_2, S_1), S_1^{\mu \nu} \right],
\end{equation}
which does not obviously cancel. However, if we schematically plug in for the amplitudes using Eq.~(\ref{treeS0}) and Eq.~(\ref{treeS1}), the necessary steps become clearer
\begin{equation}
    \frac{e^{i q \cdot \omega}}{4} \int \cancel{D}l\,\frac{i \mathcal{N}^{(1, 0)} \mathcal{N}^{(1, 1)}_{\alpha}(\bar{u}_1,\bar{u}_2 )}{l^2 (q - l)^2} (2 l_{\beta} - q_{\beta}) \left[S_1^{\alpha \beta}, S_1^{\mu \nu}\right]. 
\end{equation}
Because the cancellation of classically singular terms occurs upon integration, we can make the same replacement for the loop momentum as in Eq.~(\ref{vecdecomp}) and apply the $\delta$-functions in $\cancel{D}l$. Doing so gives us the uniformly classical spin kick
\begin{align}\label{dsnloamps}
    \Delta^{(2, 1)}S^{\mu \nu}_1 = \, & \frac{i}{2} \int \cancel{D}q \, e^{- i q \cdot b_{\text{cov}}}  \left\{ \left[\text{Re} \mathcal{A}^{(2, 1)}(q, \Bar{u}_1, \Bar{u}_2, S_1), S^{\mu \nu}_1 \right] \right. \nonumber\\
    & \left. \null+ \int \cancel{D}l \left\{\frac{ (-q^2)}{4 (y^2 - 1)} \frac{\mathcal{A}^{(1, 0)}(l) \mathcal{N}^{(1, 1)}_{\alpha}(\Bar{u}_1, \Bar{u}_2)}{(q - l)^2} \left(\frac{\check{\bar{u}}_{1\, \beta}}{\bar{m}_1} - \frac{\check{\bar{u}}_{2\, \beta }}{\bar{m}_2}\right) \left[S^{\alpha \beta}_1, S^{\mu \nu}_1 \right] \right.\right. \nonumber\\
    &\left.\left. \hspace{1.4cm} \null- \frac{i}{2} \left[\mathcal{A}^{(1, 1)}(q- l, \Bar{u}_1, \Bar{u}_2, S_1), \left[\mathcal{A}^{(1, 1)}(l, \Bar{u}_1, \Bar{u}_2, S_1), S^{\mu \nu}_1\right] \right] \right.\right.\nonumber \\
    & \left.\left. \hspace{1.4cm} \null+ \frac{i}{4}  \left[\left[\mathcal{A}^{(1, 1)}(q-l,\Bar{u}_1, \Bar{u}_2, S_1), \mathcal{A}^{(1, 1)}(l, \Bar{u}_1, \Bar{u}_2, S_1)\right], S^{\mu \nu}_1 \right] \right\}\right\} . 
\end{align}
Note that the last line in Eq.~(\ref{dsnloamps}) evaluates to zero due to the horizontal flip symmetry of the cut, i.e. $l \rightarrow q - l$. Once again, we systematically replace our amplitudes with eikonal phases by shifting $q \rightarrow q + l$ and replacing explicit factors of $q, l$ with $i \frac{\partial}{\partial b}$,
\begin{align}\label{dsnloeik}
    \Delta^{(2, 1)}S^{\mu \nu}_{1} =& \frac{i}{2} \left[\delta_{\text{cov}}^{(2, 1)}, S^{\mu \nu}_1 \right] + \frac{1}{4}  \left[\delta_{\text{cov}}^{(1, 1)}, \left[\delta_{\text{cov}}^{(1, 1)}, S^{\mu \nu}_1\right] \right] \nonumber \\
    &  \null+ \frac{i \mathcal{N}^{(1, 1)}_{\alpha}(\Bar{u}_1, \Bar{u}_2)}{32 \pi \bar{m}_1 \bar{m}_2 (y^2 - 1)^{3/2}} \left(\frac{\check{\bar{u}}_{1\, \beta}}{\bar{m}_1} - \frac{\check{\bar{u}}_{2\, \beta }}{\bar{m}_2}\right) \left[S^{\alpha \beta}_1, S^{\mu \nu}_1 \right] \frac{\partial \delta_{\text{cov}}^{(1, 0)} }{\partial b^{\gamma}} \frac{b^{\gamma}_{\text{cov}}}{|b_{\text{cov}}|^2} .
\end{align} 
We verify Eq.~(\ref{dsnloeik}) by plugging in the appropriate amplitudes from Eq.~(\ref{lagdens})
\begin{align}
    &\Delta^{(2, 1)}S^{\mu \nu}_{1} =
%%%%% begin : DeltaS[2,1]
   \frac{\pi S^{\rho [\nu}_1 \eta^{\mu] \sigma}}{|b_{\text{cov}}|^3 (y^2 - 1)^{3/2} \, \Bar{m}_1^2 \, \Bar{m}_2} \left[b_{\text{cov}\,[\sigma}\check{\Bar{u}}_{1 \, \rho]} \left(\Bar{m}_1(D_1 - C_1) - \Bar{m}_2(C_1^2 - D_1 - C_1 D_1)\right) \right. \nonumber \\
    & \left. \hspace{6.1cm} \null+ b_{\text{cov} \, [\sigma}\check{\Bar{u}}_{2 \, \rho]}  y \left( \Bar{m}_1 (D_1 + C_1) - \Bar{m}_2 D_1(C_1 - D_1 -3) \right) \right] \nonumber \\
    &\hspace{2.0cm} \null - \frac{2}{|b_{\text{cov}}|^4 (y^2 - 1) \, \Bar{m}_1^2} \left[2 b_{\text{cov} \, \rho} S_1^{\rho \sigma}\left(C_1 \Bar{u}_{2\,\sigma} - D_1 y \Bar{u}_{1\,\sigma}\right) \left(C_1 b_{\text{cov}}^{[\mu} \Bar{u}_2^{\nu]} \null- y D_1 b_{\text{cov}}^{[\mu} \Bar{u}_1^{\nu]}\right) \right. \nonumber \\ 
    &\left. \hspace{5.6cm} \null -b_{\text{cov}\, \rho}  S_1^{\rho [\nu}b_{\text{cov}}^{\mu]} (C_1^2 - D_1(2C_1 - D_1)y^2)\right] \nonumber \\
    & \null- \frac{2}{|b_{\text{cov}}|^2 (y^2 - 1)^3 \, \Bar{m}_1^2 \, \Bar{m}_2} \left[\Bar{m}_2 \left(y^2 (C_1 - D_1)^2 S_1^{ \rho [\nu} \check{\Bar{u}}^{\mu]}_1\check{\Bar{u}}_{1\, \rho}  + (C_1 - D_1 y^2)^2   S_1^{\rho [\nu} \check{\Bar{u}}^{\mu]}_2\check{\Bar{u}}_{2\, \rho}\right) \right. \nonumber \\ 
    & \left. \hspace{3.5cm} \null+  S_{1}^{\rho [\nu}\check{\Bar{u}}^{\mu]}_1 \check{\Bar{u}}_{2\, \rho} \left(\Bar{m}_1 (C_1 - D_1)y^2 + y \Bar{m}_2 (C_1 - y^2 D_1 )(C_1 - D_1 + 1) \right) \right. \nonumber\\
    &\left. \hspace{3.5cm} \null- S_{1}^{\rho [\nu} \check{\Bar{u}}^{\mu]}_2 \check{\Bar{u}}_{1\, \rho} (\Bar{m}_1 (C_1 - D_1)y^2 - y \Bar{m}_2 (C_1 - y^2 D_1 )(C_1 - D_1 - 1))\right]
%%%%% end : DeltaS[2,1]
, 
\end{align}
which matches the results in Ref.~\cite{Bern:2023ity} taking in to account the difference in an overall factor of $1/2$ due to our definition of the spin operator and our impact parameters pointing in opposite directions. Taking into account the fact that we use the spin tensor and the covariant impact parameter, we see that the top line of Eq.~(\ref{dsnloeik}) has a similar form to Eq.~(\ref{DSLformds}) not including the $\mathcal{D}_L(f, g)$ terms. Because we have verified Eq.~(\ref{dsnloeik}) using the result in Ref.~\cite{Bern:2023ity}, we conclude that we have derived an eikonal formula for the one loop, linear-in-spin spin kick that takes into account the contributions of the $\mathcal{D}_{L}(f, g)$ terms in Ref.~\cite{Bern:2023ity}. 

\section{Conclusion and Discussion}
We derived covariant eikonal formulae relating the eikonal phase to the momentum impulse and spin kick to linear order in spin and next to leading order in the coupling constant. To do so we used the KMOC formalism \cite{Kosower:2018adc}, which directly gives physical observables in terms of scattering amplitudes in impact parameter space, which is connected to the eikonal phase. This is the first next-to-leading order spin dependent calculation using the KMOC formalism and general spin QFT.
To simplify the evaluation, we used the non-transverse property of unconstrained massive general spin fields \cite{Bern:2023ity} to construct the completeness relation Eq.~(\ref{complrel}) for the polarization tensors. This non-transverse property also allows us to construct SSC-violating spin tensors, as in the worldline and field theory calculations of \cite{Bern:2023ity}. The choice of special kinematics \cite{Herrmann:2021tct, Parra-Martinez:2020dzs} simplified the cancellation of classically singular terms in the integrand and facilitated the appearance of terms proportional to external momentum derivatives. We verified our results by comparing to the results in Ref.~\cite{Bern:2023ity}, which were based on worldline and field theory matched to a general spin EFT. Based on this match, we also conclude that our eikonal formulae, Eqs.~(\ref{dpeik}) and  (\ref{dsnloeik}), capture the effects of the $\mathcal{D}_{L}(f, g)$ terms in Eqs.~(\ref{DSLformdp}) and (\ref{DSLformds}). 

%Being able to express the real kernel and the imaginary part of the one loop amplitude explicitly using special kinematics was crucial to manifesting the cancellation of classically singular terms in the integrand; it also facilitated the appearance of terms proportional to external momentum derivatives. 

%By virtue of our match with the existing literature, we have also shown that using non-transverse polarization states for massive fields in conjunction with the use of coherent spin states is an appropriate option for calculating the classical limit of spinning observables from an unconstrained massive general spin field. 

There are obvious future directions, including extending this work to higher order in spin and coupling constant. The eikonal formulae, based on rest-frame spin introduced in Ref.~\cite{Bern:2020buy}, have been explicitly verified to quadratic order in spin \cite{Kosmopoulos:2021zoq} and next to leading order in the coupling.  Though our covariant construction is different, extending this work to quadratic-in-spin order (and beyond) will provide insight to the relationship between the eikonal phase and spinning scattering observables. It would, of course, be very interesting to extend the derivation  of the eikonal formulae to higher orders in the coupling as well. 
%As shown in Refs.~\cite{DiVecchia:2021bdo, DiVecchia:2022piu, DiVecchia:2023frv}, at two loops the eikonal phase starts to have imaginary components. 
%At two loops the structure of the imaginary part of the amplitude is more complicated and includes one-loop contributions, which will complicate the explicit cancellation of classically singular terms. 

A more thorough investigation of the relation between the eikonal formulae for the impulse and spin kick presented here and those in Refs.~\cite{Bern:2023ity} would be very interesting. It would also be interesting to use the procedure presented here to incorporate spin in the calculation of other observables such as the direct waveform calculations \cite{Aoude:2023dui, Brandhuber:2023hhl, DeAngelis:2023lvf, Herderschee:2023fxh, Adamo:2022qci}, absorption effects \cite{Jones:2023ugm}, or even to observables in the three-body case \cite{Jones:2022aji}.

\section*{Acknowledgements}

We thank Zvi Bern,  Callum Jones, Lukas Lindwasser, Richard Myers, and Trevor Scheopner for numerous insightful discussions and helpful guidance throughout this work. We also thank Rafael Aoude for informative discussions on spin coherent states. We thank Zvi Bern, Dimitrios Kosmopoulos, Andres Luna, Radu Roiban, Trevor Scheopner, Fei Teng, and Justin Vines for sharing the results of Ref.~\cite{Bern:2023ity}, which we used to verify our results. This work
was supported in part by the U.S. Department of Energy (DOE) under Award Number
DE-SC0009937. We are also grateful for support from the Mani L. Bhaumik Institute for Theoretical Physics.

\newpage

\appendix
\section{The Spin Operator}\label{spinopsec}

We consider the operator promotion of angular momentum, as done in Appendix B of Ref.~\cite{Maybee:2019jus}
\begin{equation}
    \mathbb{J}^{\mu \nu} = \int d^3 x \, M^{0 \mu \nu} = \int d^3 x \left( x^{\mu} T^{0 \nu} - x^{\nu} T^{0 \mu}\right),
\end{equation}
where $M^{\alpha \mu \nu}$ is the conserved Noether current associated with Lorentz symmetry and $T^{\mu \nu}$ is the Belinfante tensor
\begin{equation}
    T^{\mu \nu} = \Theta^{\mu \nu} - \frac{i}{2} \partial_{\rho} \left( \frac{\partial \mathcal{L}_0}{\partial(\partial_{\rho} \Phi_s)} \cdot  \mathit{M}^{\mu \nu} \cdot \Phi_s - \frac{\partial \mathcal{L}_0}{\partial(\partial_{\mu} \Phi_s)} \cdot \mathit{M}^{\rho \nu} \cdot \Phi_s - \frac{\partial \mathcal{L}_0}{\partial(\partial_{\nu} \Phi_s)} \cdot  \mathit{M}^{\rho \mu} \cdot  \Phi_s\right),
\end{equation}
where $\Theta^{\mu \nu}$ is the stress-energy tensor, $\partial_{\rho} = \frac{\partial}{\partial x^{\rho}}$, $(\mathit{M}^{\mu \nu})^{\alpha(s)}_{\,\,\,\,\,\beta(s)}$ is the Lorentz generator, the dots surrounding the Lorentz generator implies summation over the representation indices, and $\Phi_s(x)$ is the complex free general spin field
\begin{equation}
    \Phi_s(x) = \sum_{a} \int d\Phi(k) \left( a_{a}(k)\epsilon_{a}(k)e^{- i k \cdot x} + b^{\dagger}_{a}(k) \epsilon^{*}_{a}(k)e^{i k \cdot x}\right),
\end{equation}
where $a$ is the little group index and $a_{a}, b^{\dagger}_{a}$ are the particle and anti-particle annihilation operators, respectively. Assuming a reasonable free kinetic term from some Lagrangian $\mathcal{L}_0$, like Eq.~(\ref{minlagrangian}) with $\alpha \rightarrow 0$, we obtain a natural separation of angular momentum in to what we call ``orbital'' and ``spin'' parts:
\begin{equation}
    \mathbb{J}^{\mu \nu} = \int d^{3}x \: \left( x^{\mu} \Theta^{0 \nu}(x) - x^{\nu} \Theta^{0 \mu}(x) + i \Pi_s(x)\cdot \mathit{M}^{\mu \nu} \cdot \Phi_s(x) \right) = \mathbb{L}^{\mu \nu} + \mathbb{S}^{\mu \nu},
\end{equation}
$\Pi_s(x) = \frac{\partial \mathcal{L}}{\partial_{0} \Phi_s}$ is the conjugate momentum to $\Phi_s$. We call the term with the Lorentz generator the spin operator, $\mathbb{S}^{\mu \nu}$. Taking the normal ordered expectation value of the spin operator gives us the following for the free theory,
\begin{equation}
    \langle{p'; a'}| \mathbb{S}^{\mu \nu} |p ; a \rangle = \langle{0}| a^{a'}(p'):\int d^{3}x \: i \: \Pi_s \, \mathit{M}^{\mu \nu} \Phi_s: a^{\dagger}_a(p) | 0 \rangle = - \frac{1}{2} \hat{\delta}_{\Phi}(p' - p) \, \epsilon^{a'}(p')\cdot\mathit{M}^{\mu \nu} \cdot \epsilon_{a}(p),  
\end{equation}
where we use the commutation relation 
\begin{equation}
    [a_i(p'),a^{\dagger}_j(p)] = 2 E_p \, \hat{\delta}^{(3)}(\Vec{p}\,' - \Vec{p}\,) \, \delta_{i, j} = \hat{\delta}_{\Phi}(p' - p) \,\delta_{i, j},
\end{equation}  
and ignore terms proportional to $b^{\dagger}$ since we are only interested in positive energy states in the classical limit. 

\newpage

\section{Polarization Tensors and their Products}\label{poltenssec}
We know that polarization tensors are not tensors in the usual sense. To boost from one arbitrary frame to another, one has to first perform a standard boost to some arbitrary reference frame followed by a boost to the final frame: Ref.~\cite{Weinberg:1995mt} provides a thorough explanation using the rest frame as a reference frame explicitly. Refs.~\cite{Guevara:2019fsj, Chung:2019duq} provide an alternative expression for the standard boost with some arbitrary reference frame $k$ without explicitly breaking covariance.
As in Ref.~\cite{Weinberg:1995mt, Chung:2019duq}, we write our spin-s polarization tensors as 
\begin{equation}
    \epsilon^{\mu(s)}(p) = D(L(p;k))^{\mu(s)}_{\, \, \, \, \, \nu(s)} \epsilon^{\nu(s)}(k),
\end{equation}
where this $D$ matrix boosts $\epsilon^{\mu(s)}$ from reference frame $k$ to arbitrary frame $p$ via some standard boost $L(p;k)$ and furnishes a representation of the homogeneous Lorentz group. For example, the scalar representation $D = 1$ and the vector representation $D = \Lambda^{\mu}_{\,\,\,\nu}$, which is what we normally think of as a standard Lorentz transformation.

While Refs.~\cite{Weinberg:1995mt, Bern:2023ity, Chung:2018kqs} have a thorough discussion on how to deal with this feature of polarization tensors, they eventually break covariance explicitly, which is inconvenient for the KMOC formalism. Ref.~\cite{Chung:2019duq} provides an alternative expression for these $D$ matrices
\begin{align}
     D(L(p;k))^{\alpha(s)}_{\,\,\,\,\, \beta(s)} &= \text{Exp}[- i \lambda(p;k) p^{\mu} k^{\nu} \mathit{M}_{\mu \nu}]^{\alpha(s)}_{\,\,\,\,\,\beta(s)}, \\
     \lambda(p;k) &= \frac{\text{arcosh}\left(\frac{p \cdot k}{m^2}\right)}{\sqrt{(p \cdot k)^2 - m^4}} \equiv \frac{\Theta}{\sqrt{\Gamma}}.
\end{align}
The use of $\Theta$ and $\Gamma$ is for later convenience and holds no significance beyond this discussion. We implicitly take $k$ to be the rest frame but do not plug in this fact explicitly. 

We now consider a small perturbation in the desired frame $p \rightarrow p + q$. We perturbatively expand the following polarization product about small $q$
\begin{align}
    \epsilon^{*}(p+q) \cdot \epsilon(p) = & \, \epsilon^{*}(p+q)|_{q \rightarrow 0} \cdot \epsilon(p) + q^{\alpha} \left(\frac{\partial}{\partial p^{\alpha}}\epsilon^{*}(p + q) \right)|_{q \rightarrow 0} \cdot \epsilon(p) \nonumber \\ 
    & \null+ \frac{q^{\beta}q^{\alpha}}{2!} \left(\frac{\partial}{\partial p^{\beta}}\frac{\partial}{\partial p^{\alpha}}\epsilon^{*}(p + q) \right)|_{q \rightarrow 0} \cdot \epsilon(p) + \cdots
\end{align}
where we suppress little group indices for clarity. 

We start with the first non-trivial order in the expansion. Note we are now taking derivatives of a matrix exponential, therefore we must keep in mind the derivative of the exponential map
\begin{equation}
    \frac{d}{dt} e^{X(t)} = \left(\frac{1 - e^{- \text{ad}_X}}{\text{ad}_X} \frac{d X(t)}{dt}\right) e^{X(t)},
\end{equation}
where
\begin{equation}
    \frac{1 - e^{- \text{ad}_X}}{\text{ad}_X} = \sum^{\infty}_{k = 0} \frac{(-1)^k}{(k+1)!} (\text{ad}_X)^k,
\end{equation}
and 
\begin{equation}
    \text{ad}_X(Y) = [X, Y]. 
\end{equation}
Using this information, we expand the first derivative to sixth order in the exponential map expansion
\begin{align}
    q^{\alpha} \left(\frac{\partial}{\partial p^{\alpha}}\epsilon(p + q) \right)|_{q \rightarrow 0} = & \left\{p_{\mu} k_{\nu} \frac{m^2 q^2}{2} \left( \frac{\lambda^3}{3!} + \frac{\lambda^5 \Gamma}{5!} + \frac{\lambda^7 \Gamma^2}{7!} + \cdots \right) \right. \nonumber\\
    & \left. \hspace{0.25cm} \null+ k_{\mu} q_{\nu}\left( \left(\lambda + \frac{\lambda^3 \Gamma}{3!} + \frac{\lambda^5 \Gamma^2}{5!} + \frac{\lambda^7 \Gamma^3}{7!} + \cdots\right) \right. \right. \nonumber\\
    &\left. \left. \hspace{2.25cm} \null- k \cdot p \left(\frac{\lambda^2}{2!} + \frac{\lambda^4 \Gamma}{4!} + \frac{\lambda^6 \Gamma^2}{6!} + \cdots \right) \right)  \right. \nonumber\\
    & \left.  \hspace{0.25cm} \null+ p_{\mu} q_{\nu} m^2 \left( \frac{\lambda^2}{2!} + \frac{\lambda^4 \Gamma}{4!} + \frac{\lambda^6 \Gamma^2}{6!} + \cdots \right) \right\} i \mathit{M}^{\mu \nu} \epsilon(p) , 
\end{align}
where we apply $k \cdot q = 0$ since $q$ is assumed to be space-like and $k$ is the rest frame, and $ p \cdot q = - q^2/2$ due to momentum conservation. Recall that the $\lambda$ carry factors of $\Gamma$; taking this into account results in equal scaling of $\Gamma$ in each series resulting in the following resummation
\begin{align}
    q^{\alpha} \left(\frac{\partial}{\partial p^{\alpha}}\epsilon(p + q) \right)|_{q \rightarrow 0} =& \left\{p_{\mu} k_{\nu} \frac{m^2 q^2}{2 \Gamma} \left( \frac{\sinh{\Theta}}{\sqrt{\Gamma}} - \lambda\right) \right. \nonumber \\
    & \left. \hspace{0.25cm} \null+k_{\mu} q_{\nu}\left( \frac{\sinh{\Theta}}{\sqrt{\Gamma}} - \frac{k \cdot p}{\Gamma} \left(\cosh{\Theta} - 1 \right) \right) \right. \nonumber \\
    & \left. \hspace{0.25cm} \null+ p_{\mu} q_{\nu} \frac{m^2}{\Gamma} \left(\cosh{\Theta} - 1 \right) \right\} i \mathit{M}^{\mu \nu} \epsilon(p). 
\end{align}
Finally, a couple of useful relations to know are $\cosh{(\text{arcosh}x)} = x$ and $\sinh{(\text{arcosh}x)} = \sqrt{x^2 - 1}$. With this in mind we arrive at the first term in the expansion
\begin{equation}
     q^{\alpha} \left(\frac{\partial}{\partial p^{\alpha}}\epsilon(p + q) \right)|_{q \rightarrow 0} = \left((k + p)_{\mu} q_{\nu} + p_{\mu} k_{\nu} \frac{q^2 (1 - m^2 \lambda)}{2 (p\cdot k - m^2)} \right)\frac{i \mathit{M}^{\mu \nu}}{p \cdot k + m^2} \epsilon(p),
\end{equation}
where we should note that the second term in the parentheses is higher order in our expansion parameter $q$, and will therefore be ignored in the future.  

We now have an answer for the first order expansion of the product of polarization tensors
\begin{equation}
    q^{\alpha} \left(\frac{\partial}{\partial p^{\alpha}}\epsilon^{*}(p + q) \right)|_{q \rightarrow 0} \cdot \epsilon(p) =  i q_{\mu} \frac{S^{\mu \nu}(p) (k_{\nu}  + p_{\nu} ) }{p \cdot k + m^2}  \equiv i q_{\mu} \omega^{\mu}.
\end{equation}
Calculating higher orders in the perturbation series streamlines. We consider the second derivative of the exponential map
\begin{equation}
    \frac{d^2}{dt^2} e^{X(t)} = \frac{d}{dt}\left(\frac{1 - e^{- \text{ad}_X}}{\text{ad}_X} \frac{d X(t)}{dt}\right) e^{X(t)} + \left(\frac{1 - e^{- \text{ad}_X}}{\text{ad}_X} \frac{d X(t)}{dt}\right)^2 e^{X(t)}. 
\end{equation}
We have just calculated the terms in the parentheses, so all we have to do is take our previous expression and simply plug in
\begin{align}
    \frac{q^{\beta}q^{\alpha}}{2!} \left(\frac{\partial}{\partial p^{\beta}}\frac{\partial}{\partial p^{\alpha}}\epsilon(p + q) \right)|_{q \rightarrow 0} =& \frac{1}{2}\left[- i k_{\mu} q_{\nu} \mathit{M}^{\mu \nu}\frac{q^2 (1 - m^2 \lambda)}{2 (p \cdot k^2 - m^4)} \right. \nonumber\\ 
    & \left. \hspace{0.6cm} \null+ \left((k + p)_{\mu} q_{\nu} \frac{i \mathit{M}^{\mu \nu}}{p \cdot k + m^2}\right. \right. \nonumber\\
    &\left.\left. \hspace{1.3cm} \null+ p_{\mu} k_{\nu} \frac{q^2 (1 - m^2 \lambda)}{2 (p\cdot k - m^2)}\frac{i \mathit{M}^{\mu \nu}}{p \cdot k + m^2} \right)^2\right]  \epsilon(p).
\end{align}
Once again, looking at the $q$-scaling of our expression, we only have one term that contributes to the leading order, which gives us the truncated second order term
\begin{align}
    \frac{q^{\beta}q^{\alpha}}{2!} \left(\frac{\partial}{\partial p^{\beta}}\frac{\partial}{\partial p^{\alpha}}\epsilon^{*}(p + q) \right)|_{q \rightarrow 0} \cdot \epsilon(p) =& q_{\mu} q_{\rho} \frac{- (k_{\nu}+p_{\nu})(k_{\sigma}+p_{\sigma})}{2 (p \cdot k + m^2)^2} \epsilon^{*}(p)\frac{1}{2}\left\{\mathit{M}^{\mu \nu},\mathit{M}^{\rho \sigma}\right\} \epsilon(p) \nonumber \\
    &= q_{\mu} q_{\rho} \frac{-  S^{\mu \nu}(p)(k_{\nu}+p_{\nu})  S^{\rho \sigma}(p)(k_{\sigma}+p_{\sigma})}{2 (p \cdot k + m^2)^2} ,
\end{align}
where we preemptively decomposed the quadratic product of Lorentz generators in to symmetric and anti-symmetric parts. We also use the identity introduced in Ref.~\cite{Bern:2020buy} relating symmetric products of Lorentz generators and products of the spin tensor. Here we begin to notice a pattern
\begin{align}
     \frac{q^{\beta}q^{\alpha}}{2!} & \left(\frac{\partial}{\partial p^{\beta}}\frac{\partial}{\partial p^{\alpha}}\epsilon^{*}(p + q) \right)|_{q \rightarrow 0} \cdot \epsilon(p) = \frac{\left(i q_{\mu} \omega^{\mu}\right) \left(i q_{\rho} \omega^{\rho}\right)}{2!}. \nonumber
\end{align}
At this point, we see the same pattern of exponentiation as seen in Refs.~\cite{Bern:2023ity, Bern:2020buy, Bern:2022kto, Kosmopoulos:2021zoq} and use the exponentiated form of the polarization tensor product Eq.~(\ref{polprodexp}). 

\newpage

\bibliography{jheprefs.bib}

\providecommand{\href}[2]{#2}\begingroup\raggedright\begin{thebibliography}{100}

\bibitem{LIGOScientific:2016aoc}
{\bf LIGO Scientific, Virgo} Collaboration, B.~P. Abbott et~al., {\it
  {Observation of Gravitational Waves from a Binary Black Hole Merger}},  {\em
  Phys. Rev. Lett.} {\bf 116} (2016), no.~6 061102,
  [\href{http://arxiv.org/abs/1602.03837}{{\tt arXiv:1602.03837}}].

\bibitem{lIGOScientific:2017vwq}
{\bf LIGO Scientific, Virgo} Collaboration, B.~P. Abbott et~al., {\it
  {GW170817: Observation of Gravitational Waves from a Binary Neutron Star
  Inspiral}},  {\em Phys. Rev. Lett.} {\bf 119} (2017), no.~16 161101,
  [\href{http://arxiv.org/abs/1710.05832}{{\tt arXiv:1710.05832}}].

\bibitem{Punturo:2010zz}
M.~Punturo et~al., {\it {The Einstein Telescope: A third-generation
  gravitational wave observatory}},  {\em Class. Quant. Grav.} {\bf 27} (2010)
  194002.

\bibitem{LISA:2017pwj}
{\bf LISA} Collaboration, P.~Amaro-Seoane et~al., {\it {Laser Interferometer
  Space Antenna}},  \href{http://arxiv.org/abs/1702.00786}{{\tt
  arXiv:1702.00786}}.

\bibitem{Reitze:2019iox}
D.~Reitze et~al., {\it {Cosmic Explorer: The U.S. Contribution to
  Gravitational-Wave Astronomy beyond LIGO}},  {\em Bull. Am. Astron. Soc.}
  {\bf 51} (2019), no.~7 035, [\href{http://arxiv.org/abs/1907.04833}{{\tt
  arXiv:1907.04833}}].

\bibitem{Kosower:2018adc}
D.~A. Kosower, B.~Maybee, and D.~O'Connell, {\it {Amplitudes, Observables, and
  Classical Scattering}},  {\em JHEP} {\bf 02} (2019) 137,
  [\href{http://arxiv.org/abs/1811.10950}{{\tt arXiv:1811.10950}}].

\bibitem{Bern:2023ity}
Z.~Bern, D.~Kosmopoulos, A.~Luna, R.~Roiban, T.~Scheopner, F.~Teng, and
  J.~Vines, {\it {Quantum Field Theory, Worldline Theory, and Spin Magnitude
  Change in Orbital Evolution}},  \href{http://arxiv.org/abs/2308.14176}{{\tt
  arXiv:2308.14176}}.

\bibitem{Mathisson:1937zz}
M.~Mathisson, {\it {Neue mechanik materieller systemes}},  {\em Acta Phys.
  Polon.} {\bf 6} (1937) 163--200.

\bibitem{Papapetrou:1951pa}
A.~Papapetrou, {\it {Spinning test particles in general relativity. 1.}},  {\em
  Proc. Roy. Soc. Lond. A} {\bf 209} (1951) 248--258.

\bibitem{Pirani:1956tn}
F.~A.~E. Pirani, {\it {On the Physical significance of the Riemann tensor}},
  {\em Acta Phys. Polon.} {\bf 15} (1956) 389--405.

\bibitem{Tulczyjew:1959}
W.~Tulczyjew, {\it {Equations of Motion of Rotating Bodies in General
  Relativity Theory}},  {\em Acta Phys. Polon.} {\bf 18} (1959) 37--55.
  [Erratum: Acta Phys. Pol. 18, 393 (1959)].

\bibitem{Barker:1970zr}
B.~M. Barker and R.~F. O'Connell, {\it {Derivation of the equations of motion
  of a gyroscope from the quantum theory of gravitation}},  {\em Phys. Rev. D}
  {\bf 2} (1970) 1428--1435.

\bibitem{Barker:1975ae}
B.~M. Barker and R.~F. O'Connell, {\it {Gravitational Two-Body Problem with
  Arbitrary Masses, Spins, and Quadrupole Moments}},  {\em Phys. Rev. D} {\bf
  12} (1975) 329--335.

\bibitem{Kidder:1992fr}
L.~E. Kidder, C.~M. Will, and A.~G. Wiseman, {\it {Spin effects in the inspiral
  of coalescing compact binaries}},  {\em Phys. Rev. D} {\bf 47} (1993), no.~10
  R4183--R4187, [\href{http://arxiv.org/abs/gr-qc/9211025}{{\tt
  gr-qc/9211025}}].

\bibitem{Kidder:1995zr}
L.~E. Kidder, {\it {Coalescing binary systems of compact objects to
  postNewtonian 5/2 order. 5. Spin effects}},  {\em Phys. Rev. D} {\bf 52}
  (1995) 821--847, [\href{http://arxiv.org/abs/gr-qc/9506022}{{\tt
  gr-qc/9506022}}].

\bibitem{Blanchet:1998vx}
L.~Blanchet, G.~Faye, and B.~Ponsot, {\it {Gravitational field and equations of
  motion of compact binaries to 5/2 postNewtonian order}},  {\em Phys. Rev. D}
  {\bf 58} (1998) 124002, [\href{http://arxiv.org/abs/gr-qc/9804079}{{\tt
  gr-qc/9804079}}].

\bibitem{Tagoshi:2000zg}
H.~Tagoshi, A.~Ohashi, and B.~J. Owen, {\it {Gravitational field and equations
  of motion of spinning compact binaries to 2.5 postNewtonian order}},  {\em
  Phys. Rev. D} {\bf 63} (2001) 044006,
  [\href{http://arxiv.org/abs/gr-qc/0010014}{{\tt gr-qc/0010014}}].

\bibitem{Porto:2005ac}
R.~A. Porto, {\it {Post-Newtonian corrections to the motion of spinning bodies
  in NRGR}},  {\em Phys. Rev. D} {\bf 73} (2006) 104031,
  [\href{http://arxiv.org/abs/gr-qc/0511061}{{\tt gr-qc/0511061}}].

\bibitem{Faye:2006gx}
G.~Faye, L.~Blanchet, and A.~Buonanno, {\it {Higher-order spin effects in the
  dynamics of compact binaries. I. Equations of motion}},  {\em Phys. Rev. D}
  {\bf 74} (2006) 104033, [\href{http://arxiv.org/abs/gr-qc/0605139}{{\tt
  gr-qc/0605139}}].

\bibitem{Blanchet:2006gy}
L.~Blanchet, A.~Buonanno, and G.~Faye, {\it {Higher-order spin effects in the
  dynamics of compact binaries. II. Radiation field}},  {\em Phys. Rev. D} {\bf
  74} (2006) 104034, [\href{http://arxiv.org/abs/gr-qc/0605140}{{\tt
  gr-qc/0605140}}]. [Erratum: Phys.Rev.D 75, 049903 (2007), Erratum: Phys.Rev.D
  81, 089901 (2010)].

\bibitem{Damour:2007nc}
T.~Damour, P.~Jaranowski, and G.~Schaefer, {\it {Hamiltonian of two spinning
  compact bodies with next-to-leading order gravitational spin-orbit
  coupling}},  {\em Phys. Rev. D} {\bf 77} (2008) 064032,
  [\href{http://arxiv.org/abs/0711.1048}{{\tt arXiv:0711.1048}}].

\bibitem{Steinhoff:2007mb}
J.~Steinhoff, S.~Hergt, and G.~Schaefer, {\it {On the next-to-leading order
  gravitational spin(1)-spin(2) dynamics}},  {\em Phys. Rev. D} {\bf 77} (2008)
  081501, [\href{http://arxiv.org/abs/0712.1716}{{\tt arXiv:0712.1716}}].

\bibitem{Levi:2008nh}
M.~Levi, {\it {Next to Leading Order gravitational Spin1-Spin2 coupling with
  Kaluza-Klein reduction}},  {\em Phys. Rev. D} {\bf 82} (2010) 064029,
  [\href{http://arxiv.org/abs/0802.1508}{{\tt arXiv:0802.1508}}].

\bibitem{Steinhoff:2008zr}
J.~Steinhoff, G.~Schaefer, and S.~Hergt, {\it {ADM canonical formalism for
  gravitating spinning objects}},  {\em Phys. Rev. D} {\bf 77} (2008) 104018,
  [\href{http://arxiv.org/abs/0805.3136}{{\tt arXiv:0805.3136}}].

\bibitem{Steinhoff:2008ji}
J.~Steinhoff, S.~Hergt, and G.~Schaefer, {\it {Spin-squared Hamiltonian of
  next-to-leading order gravitational interaction}},  {\em Phys. Rev. D} {\bf
  78} (2008) 101503, [\href{http://arxiv.org/abs/0809.2200}{{\tt
  arXiv:0809.2200}}].

\bibitem{Marsat:2012fn}
S.~Marsat, A.~Bohe, G.~Faye, and L.~Blanchet, {\it {Next-to-next-to-leading
  order spin-orbit effects in the equations of motion of compact binary
  systems}},  {\em Class. Quant. Grav.} {\bf 30} (2013) 055007,
  [\href{http://arxiv.org/abs/1210.4143}{{\tt arXiv:1210.4143}}].

\bibitem{Hergt:2010pa}
S.~Hergt, J.~Steinhoff, and G.~Schaefer, {\it {Reduced Hamiltonian for
  next-to-leading order Spin-Squared Dynamics of General Compact Binaries}},
  {\em Class. Quant. Grav.} {\bf 27} (2010) 135007,
  [\href{http://arxiv.org/abs/1002.2093}{{\tt arXiv:1002.2093}}].

\bibitem{Porto:2010tr}
R.~A. Porto, {\it {Next to leading order spin-orbit effects in the motion of
  inspiralling compact binaries}},  {\em Class. Quant. Grav.} {\bf 27} (2010)
  205001, [\href{http://arxiv.org/abs/1005.5730}{{\tt arXiv:1005.5730}}].

\bibitem{Levi:2010zu}
M.~Levi, {\it {Next to Leading Order gravitational Spin-Orbit coupling in an
  Effective Field Theory approach}},  {\em Phys. Rev. D} {\bf 82} (2010)
  104004, [\href{http://arxiv.org/abs/1006.4139}{{\tt arXiv:1006.4139}}].

\bibitem{Porto:2010zg}
R.~A. Porto, A.~Ross, and I.~Z. Rothstein, {\it {Spin induced multipole moments
  for the gravitational wave flux from binary inspirals to third Post-Newtonian
  order}},  {\em JCAP} {\bf 03} (2011) 009,
  [\href{http://arxiv.org/abs/1007.1312}{{\tt arXiv:1007.1312}}].

\bibitem{Levi:2011eq}
M.~Levi, {\it {Binary dynamics from spin1-spin2 coupling at fourth
  post-Newtonian order}},  {\em Phys. Rev. D} {\bf 85} (2012) 064043,
  [\href{http://arxiv.org/abs/1107.4322}{{\tt arXiv:1107.4322}}].

\bibitem{Porto:2012as}
R.~A. Porto, A.~Ross, and I.~Z. Rothstein, {\it {Spin induced multipole moments
  for the gravitational wave amplitude from binary inspirals to 2.5
  Post-Newtonian order}},  {\em JCAP} {\bf 09} (2012) 028,
  [\href{http://arxiv.org/abs/1203.2962}{{\tt arXiv:1203.2962}}].

\bibitem{Hergt:2012zx}
S.~Hergt, J.~Steinhoff, and G.~Schaefer, {\it {On the comparison of results
  regarding the post-Newtonian approximate treatment of the dynamics of
  extended spinning compact binaries}},  {\em J. Phys. Conf. Ser.} {\bf 484}
  (2014) 012018, [\href{http://arxiv.org/abs/1205.4530}{{\tt
  arXiv:1205.4530}}].

\bibitem{Bohe:2012mr}
A.~Bohe, S.~Marsat, G.~Faye, and L.~Blanchet, {\it {Next-to-next-to-leading
  order spin-orbit effects in the near-zone metric and precession equations of
  compact binaries}},  {\em Class. Quant. Grav.} {\bf 30} (2013) 075017,
  [\href{http://arxiv.org/abs/1212.5520}{{\tt arXiv:1212.5520}}].

\bibitem{Hartung:2013dza}
J.~Hartung, J.~Steinhoff, and G.~Schafer, {\it {Next-to-next-to-leading order
  post-Newtonian linear-in-spin binary Hamiltonians}},  {\em Annalen Phys.}
  {\bf 525} (2013) 359--394, [\href{http://arxiv.org/abs/1302.6723}{{\tt
  arXiv:1302.6723}}].

\bibitem{Marsat:2013wwa}
S.~Marsat, L.~Blanchet, A.~Bohe, and G.~Faye, {\it {Gravitational waves from
  spinning compact object binaries: New post-Newtonian results}},  12, 2013.
\newblock \href{http://arxiv.org/abs/1312.5375}{{\tt arXiv:1312.5375}}.

\bibitem{Levi:2014gsa}
M.~Levi and J.~Steinhoff, {\it {Leading order finite size effects with spins
  for inspiralling compact binaries}},  {\em JHEP} {\bf 06} (2015) 059,
  [\href{http://arxiv.org/abs/1410.2601}{{\tt arXiv:1410.2601}}].

\bibitem{Vaidya:2014kza}
V.~Vaidya, {\it {Gravitational spin Hamiltonians from the S matrix}},  {\em
  Phys. Rev. D} {\bf 91} (2015), no.~2 024017,
  [\href{http://arxiv.org/abs/1410.5348}{{\tt arXiv:1410.5348}}].

\bibitem{Bohe:2015ana}
A.~Boh\'e, G.~Faye, S.~Marsat, and E.~K. Porter, {\it {Quadratic-in-spin
  effects in the orbital dynamics and gravitational-wave energy flux of compact
  binaries at the 3PN order}},  {\em Class. Quant. Grav.} {\bf 32} (2015),
  no.~19 195010, [\href{http://arxiv.org/abs/1501.01529}{{\tt
  arXiv:1501.01529}}].

\bibitem{Bini:2017pee}
D.~Bini, A.~Geralico, and J.~Vines, {\it {Hyperbolic scattering of spinning
  particles by a Kerr black hole}},  {\em Phys. Rev. D} {\bf 96} (2017), no.~8
  084044, [\href{http://arxiv.org/abs/1707.09814}{{\tt arXiv:1707.09814}}].

\bibitem{Siemonsen:2017yux}
N.~Siemonsen, J.~Steinhoff, and J.~Vines, {\it {Gravitational waves from
  spinning binary black holes at the leading post-Newtonian orders at all
  orders in spin}},  {\em Phys. Rev. D} {\bf 97} (2018), no.~12 124046,
  [\href{http://arxiv.org/abs/1712.08603}{{\tt arXiv:1712.08603}}].

\bibitem{Porto:2006bt}
R.~A. Porto and I.~Z. Rothstein, {\it {The Hyperfine Einstein-Infeld-Hoffmann
  potential}},  {\em Phys. Rev. Lett.} {\bf 97} (2006) 021101,
  [\href{http://arxiv.org/abs/gr-qc/0604099}{{\tt gr-qc/0604099}}].

\bibitem{Porto:2007tt}
R.~A. Porto and I.~Z. Rothstein, {\it {Comment on `On the next-to-leading order
  gravitational spin(1) - spin(2) dynamics' by J. Steinhoff et al}},
  \href{http://arxiv.org/abs/0712.2032}{{\tt arXiv:0712.2032}}.

\bibitem{Porto:2008tb}
R.~A. Porto and I.~Z. Rothstein, {\it {Spin(1)Spin(2) Effects in the Motion of
  Inspiralling Compact Binaries at Third Order in the Post-Newtonian
  Expansion}},  {\em Phys. Rev. D} {\bf 78} (2008) 044012,
  [\href{http://arxiv.org/abs/0802.0720}{{\tt arXiv:0802.0720}}]. [Erratum:
  Phys.Rev.D 81, 029904 (2010)].

\bibitem{Porto:2008jj}
R.~A. Porto and I.~Z. Rothstein, {\it {Next to Leading Order Spin(1)Spin(1)
  Effects in the Motion of Inspiralling Compact Binaries}},  {\em Phys. Rev. D}
  {\bf 78} (2008) 044013, [\href{http://arxiv.org/abs/0804.0260}{{\tt
  arXiv:0804.0260}}]. [Erratum: Phys.Rev.D 81, 029905 (2010)].

\bibitem{Levi:2014sba}
M.~Levi and J.~Steinhoff, {\it {Equivalence of ADM Hamiltonian and Effective
  Field Theory approaches at next-to-next-to-leading order spin1-spin2 coupling
  of binary inspirals}},  {\em JCAP} {\bf 12} (2014) 003,
  [\href{http://arxiv.org/abs/1408.5762}{{\tt arXiv:1408.5762}}].

\bibitem{Levi:2015msa}
M.~Levi and J.~Steinhoff, {\it {Spinning gravitating objects in the effective
  field theory in the post-Newtonian scheme}},  {\em JHEP} {\bf 09} (2015) 219,
  [\href{http://arxiv.org/abs/1501.04956}{{\tt arXiv:1501.04956}}].

\bibitem{Levi:2015uxa}
M.~Levi and J.~Steinhoff, {\it {Next-to-next-to-leading order gravitational
  spin-orbit coupling via the effective field theory for spinning objects in
  the post-Newtonian scheme}},  {\em JCAP} {\bf 01} (2016) 011,
  [\href{http://arxiv.org/abs/1506.05056}{{\tt arXiv:1506.05056}}].

\bibitem{Levi:2015ixa}
M.~Levi and J.~Steinhoff, {\it {Next-to-next-to-leading order gravitational
  spin-squared potential via the effective field theory for spinning objects in
  the post-Newtonian scheme}},  {\em JCAP} {\bf 01} (2016) 008,
  [\href{http://arxiv.org/abs/1506.05794}{{\tt arXiv:1506.05794}}].

\bibitem{Levi:2016ofk}
M.~Levi and J.~Steinhoff, {\it {Complete conservative dynamics for inspiralling
  compact binaries with spins at the fourth post-Newtonian order}},  {\em JCAP}
  {\bf 09} (2021) 029, [\href{http://arxiv.org/abs/1607.04252}{{\tt
  arXiv:1607.04252}}].

\bibitem{Levi:2019kgk}
M.~Levi, S.~Mougiakakos, and M.~Vieira, {\it {Gravitational cubic-in-spin
  interaction at the next-to-leading post-Newtonian order}},  {\em JHEP} {\bf
  01} (2021) 036, [\href{http://arxiv.org/abs/1912.06276}{{\tt
  arXiv:1912.06276}}].

\bibitem{Levi:2020lfn}
M.~Levi and F.~Teng, {\it {NLO gravitational quartic-in-spin interaction}},
  {\em JHEP} {\bf 01} (2021) 066, [\href{http://arxiv.org/abs/2008.12280}{{\tt
  arXiv:2008.12280}}].

\bibitem{Levi:2020kvb}
M.~Levi, A.~J. Mcleod, and M.~Von~Hippel, {\it {N$^{3}$LO gravitational
  spin-orbit coupling at order G$^{4}$}},  {\em JHEP} {\bf 07} (2021) 115,
  [\href{http://arxiv.org/abs/2003.02827}{{\tt arXiv:2003.02827}}].

\bibitem{Levi:2020uwu}
M.~Levi, A.~J. Mcleod, and M.~Von~Hippel, {\it {N$^{3}$LO gravitational
  quadratic-in-spin interactions at G$^{4}$}},  {\em JHEP} {\bf 07} (2021) 116,
  [\href{http://arxiv.org/abs/2003.07890}{{\tt arXiv:2003.07890}}].

\bibitem{Kim:2021rfj}
J.-W. Kim, M.~Levi, and Z.~Yin, {\it {Quadratic-in-spin interactions at fifth
  post-Newtonian order probe new physics}},  {\em Phys. Lett. B} {\bf 834}
  (2022) 137410, [\href{http://arxiv.org/abs/2112.01509}{{\tt
  arXiv:2112.01509}}].

\bibitem{Maia:2017gxn}
N.~T. Maia, C.~R. Galley, A.~K. Leibovich, and R.~A. Porto, {\it {Radiation
  reaction for spinning bodies in effective field theory I: Spin-orbit
  effects}},  {\em Phys. Rev. D} {\bf 96} (2017), no.~8 084064,
  [\href{http://arxiv.org/abs/1705.07934}{{\tt arXiv:1705.07934}}].

\bibitem{Maia:2017yok}
N.~T. Maia, C.~R. Galley, A.~K. Leibovich, and R.~A. Porto, {\it {Radiation
  reaction for spinning bodies in effective field theory II: Spin-spin
  effects}},  {\em Phys. Rev. D} {\bf 96} (2017), no.~8 084065,
  [\href{http://arxiv.org/abs/1705.07938}{{\tt arXiv:1705.07938}}].

\bibitem{Cho:2021mqw}
G.~Cho, B.~Pardo, and R.~A. Porto, {\it {Gravitational radiation from
  inspiralling compact objects: Spin-spin effects completed at the
  next-to-leading post-Newtonian order}},  {\em Phys. Rev. D} {\bf 104} (2021),
  no.~2 024037, [\href{http://arxiv.org/abs/2103.14612}{{\tt
  arXiv:2103.14612}}].

\bibitem{Cho:2022syn}
G.~Cho, R.~A. Porto, and Z.~Yang, {\it {Gravitational radiation from
  inspiralling compact objects: Spin effects to the fourth post-Newtonian
  order}},  {\em Phys. Rev. D} {\bf 106} (2022), no.~10 L101501,
  [\href{http://arxiv.org/abs/2201.05138}{{\tt arXiv:2201.05138}}].

\bibitem{Kim:2022pou}
J.-W. Kim, M.~Levi, and Z.~Yin, {\it {N$^{3}$LO spin-orbit interaction via the
  EFT of spinning gravitating objects}},  {\em JHEP} {\bf 05} (2023) 184,
  [\href{http://arxiv.org/abs/2208.14949}{{\tt arXiv:2208.14949}}].

\bibitem{Mandal:2022nty}
M.~K. Mandal, P.~Mastrolia, R.~Patil, and J.~Steinhoff, {\it {Gravitational
  spin-orbit Hamiltonian at NNNLO in the post-Newtonian framework}},  {\em
  JHEP} {\bf 03} (2023) 130, [\href{http://arxiv.org/abs/2209.00611}{{\tt
  arXiv:2209.00611}}].

\bibitem{Kim:2022bwv}
J.-W. Kim, M.~Levi, and Z.~Yin, {\it {N$^{3}$LO quadratic-in-spin interactions
  for generic compact binaries}},  {\em JHEP} {\bf 03} (2023) 098,
  [\href{http://arxiv.org/abs/2209.09235}{{\tt arXiv:2209.09235}}].

\bibitem{Mandal:2022ufb}
M.~K. Mandal, P.~Mastrolia, R.~Patil, and J.~Steinhoff, {\it {Gravitational
  quadratic-in-spin Hamiltonian at NNNLO in the post-Newtonian framework}},
  {\em JHEP} {\bf 07} (2023) 128, [\href{http://arxiv.org/abs/2210.09176}{{\tt
  arXiv:2210.09176}}].

\bibitem{Levi:2022dqm}
M.~Levi, R.~Morales, and Z.~Yin, {\it {From the EFT of spinning gravitating
  objects to Poincar\'e and gauge invariance at the 4.5PN precision frontier}},
   {\em JHEP} {\bf 09} (2023) 090, [\href{http://arxiv.org/abs/2210.17538}{{\tt
  arXiv:2210.17538}}].

\bibitem{Levi:2022rrq}
M.~Levi and Z.~Yin, {\it {Completing the fifth PN precision frontier via the
  EFT of spinning gravitating objects}},  {\em JHEP} {\bf 04} (2023) 079,
  [\href{http://arxiv.org/abs/2211.14018}{{\tt arXiv:2211.14018}}].

\bibitem{Bini:2017xzy}
D.~Bini and T.~Damour, {\it {Gravitational spin-orbit coupling in binary
  systems, post-Minkowskian approximation and effective one-body theory}},
  {\em Phys. Rev. D} {\bf 96} (2017), no.~10 104038,
  [\href{http://arxiv.org/abs/1709.00590}{{\tt arXiv:1709.00590}}].

\bibitem{Bini:2018ywr}
D.~Bini and T.~Damour, {\it {Gravitational spin-orbit coupling in binary
  systems at the second post-Minkowskian approximation}},  {\em Phys. Rev. D}
  {\bf 98} (2018), no.~4 044036, [\href{http://arxiv.org/abs/1805.10809}{{\tt
  arXiv:1805.10809}}].

\bibitem{Maybee:2019jus}
B.~Maybee, D.~O'Connell, and J.~Vines, {\it {Observables and amplitudes for
  spinning particles and black holes}},  {\em JHEP} {\bf 12} (2019) 156,
  [\href{http://arxiv.org/abs/1906.09260}{{\tt arXiv:1906.09260}}].

\bibitem{Guevara:2019fsj}
A.~Guevara, A.~Ochirov, and J.~Vines, {\it {Black-hole scattering with general
  spin directions from minimal-coupling amplitudes}},  {\em Phys. Rev. D} {\bf
  100} (2019), no.~10 104024, [\href{http://arxiv.org/abs/1906.10071}{{\tt
  arXiv:1906.10071}}].

\bibitem{Chung:2020rrz}
M.-Z. Chung, Y.-t. Huang, J.-W. Kim, and S.~Lee, {\it {Complete Hamiltonian for
  spinning binary systems at first post-Minkowskian order}},  {\em JHEP} {\bf
  05} (2020) 105, [\href{http://arxiv.org/abs/2003.06600}{{\tt
  arXiv:2003.06600}}].

\bibitem{Guevara:2017csg}
A.~Guevara, {\it {Holomorphic Classical Limit for Spin Effects in Gravitational
  and Electromagnetic Scattering}},  {\em JHEP} {\bf 04} (2019) 033,
  [\href{http://arxiv.org/abs/1706.02314}{{\tt arXiv:1706.02314}}].

\bibitem{Vines:2018gqi}
J.~Vines, J.~Steinhoff, and A.~Buonanno, {\it {Spinning-black-hole scattering
  and the test-black-hole limit at second post-Minkowskian order}},  {\em Phys.
  Rev. D} {\bf 99} (2019), no.~6 064054,
  [\href{http://arxiv.org/abs/1812.00956}{{\tt arXiv:1812.00956}}].

\bibitem{Damgaard:2019lfh}
P.~H. Damgaard, K.~Haddad, and A.~Helset, {\it {Heavy Black Hole Effective
  Theory}},  {\em JHEP} {\bf 11} (2019) 070,
  [\href{http://arxiv.org/abs/1908.10308}{{\tt arXiv:1908.10308}}].

\bibitem{Aoude:2020onz}
R.~Aoude, K.~Haddad, and A.~Helset, {\it {On-shell heavy particle effective
  theories}},  {\em JHEP} {\bf 05} (2020) 051,
  [\href{http://arxiv.org/abs/2001.09164}{{\tt arXiv:2001.09164}}].

\bibitem{Vines:2017hyw}
J.~Vines, {\it {Scattering of two spinning black holes in post-Minkowskian
  gravity, to all orders in spin, and effective-one-body mappings}},  {\em
  Class. Quant. Grav.} {\bf 35} (2018), no.~8 084002,
  [\href{http://arxiv.org/abs/1709.06016}{{\tt arXiv:1709.06016}}].

\bibitem{Guevara:2018wpp}
A.~Guevara, A.~Ochirov, and J.~Vines, {\it {Scattering of Spinning Black Holes
  from Exponentiated Soft Factors}},  {\em JHEP} {\bf 09} (2019) 056,
  [\href{http://arxiv.org/abs/1812.06895}{{\tt arXiv:1812.06895}}].

\bibitem{Chung:2018kqs}
M.-Z. Chung, Y.-T. Huang, J.-W. Kim, and S.~Lee, {\it {The simplest massive
  S-matrix: from minimal coupling to Black Holes}},  {\em JHEP} {\bf 04} (2019)
  156, [\href{http://arxiv.org/abs/1812.08752}{{\tt arXiv:1812.08752}}].

\bibitem{Chung:2019duq}
M.-Z. Chung, Y.-T. Huang, and J.-W. Kim, {\it {Classical potential for general
  spinning bodies}},  {\em JHEP} {\bf 09} (2020) 074,
  [\href{http://arxiv.org/abs/1908.08463}{{\tt arXiv:1908.08463}}].

\bibitem{Bern:2020buy}
Z.~Bern, A.~Luna, R.~Roiban, C.-H. Shen, and M.~Zeng, {\it {Spinning black hole
  binary dynamics, scattering amplitudes, and effective field theory}},  {\em
  Phys. Rev. D} {\bf 104} (2021), no.~6 065014,
  [\href{http://arxiv.org/abs/2005.03071}{{\tt arXiv:2005.03071}}].

\bibitem{Kosmopoulos:2021zoq}
D.~Kosmopoulos and A.~Luna, {\it {Quadratic-in-spin Hamiltonian at $
  \mathcal{O} $(G$^{2}$) from scattering amplitudes}},  {\em JHEP} {\bf 07}
  (2021) 037, [\href{http://arxiv.org/abs/2102.10137}{{\tt arXiv:2102.10137}}].

\bibitem{Liu:2021zxr}
Z.~Liu, R.~A. Porto, and Z.~Yang, {\it {Spin Effects in the Effective Field
  Theory Approach to Post-Minkowskian Conservative Dynamics}},  {\em JHEP} {\bf
  06} (2021) 012, [\href{http://arxiv.org/abs/2102.10059}{{\tt
  arXiv:2102.10059}}].

\bibitem{Aoude:2021oqj}
R.~Aoude and A.~Ochirov, {\it {Classical observables from coherent-spin
  amplitudes}},  {\em JHEP} {\bf 10} (2021) 008,
  [\href{http://arxiv.org/abs/2108.01649}{{\tt arXiv:2108.01649}}].

\bibitem{Jakobsen:2021lvp}
G.~U. Jakobsen, G.~Mogull, J.~Plefka, and J.~Steinhoff, {\it {Gravitational
  Bremsstrahlung and Hidden Supersymmetry of Spinning Bodies}},  {\em Phys.
  Rev. Lett.} {\bf 128} (2022), no.~1 011101,
  [\href{http://arxiv.org/abs/2106.10256}{{\tt arXiv:2106.10256}}].

\bibitem{Jakobsen:2021zvh}
G.~U. Jakobsen, G.~Mogull, J.~Plefka, and J.~Steinhoff, {\it {SUSY in the sky
  with gravitons}},  {\em JHEP} {\bf 01} (2022) 027,
  [\href{http://arxiv.org/abs/2109.04465}{{\tt arXiv:2109.04465}}].

\bibitem{Chen:2021kxt}
W.-M. Chen, M.-Z. Chung, Y.-t. Huang, and J.-W. Kim, {\it {The 2PM Hamiltonian
  for binary Kerr to quartic in spin}},  {\em JHEP} {\bf 08} (2022) 148,
  [\href{http://arxiv.org/abs/2111.13639}{{\tt arXiv:2111.13639}}].

\bibitem{Chen:2022clh}
W.-M. Chen, M.-Z. Chung, Y.-t. Huang, and J.-W. Kim, {\it {Gravitational
  Faraday effect from on-shell amplitudes}},  {\em JHEP} {\bf 12} (2022) 058,
  [\href{http://arxiv.org/abs/2205.07305}{{\tt arXiv:2205.07305}}].

\bibitem{Cristofoli:2021jas}
A.~Cristofoli, R.~Gonzo, N.~Moynihan, D.~O'Connell, A.~Ross, M.~Sergola, and
  C.~D. White, {\it {The Uncertainty Principle and Classical Amplitudes}},
  \href{http://arxiv.org/abs/2112.07556}{{\tt arXiv:2112.07556}}.

\bibitem{Chiodaroli:2021eug}
M.~Chiodaroli, H.~Johansson, and P.~Pichini, {\it {Compton black-hole
  scattering for s \ensuremath{\leq} 5/2}},  {\em JHEP} {\bf 02} (2022) 156,
  [\href{http://arxiv.org/abs/2107.14779}{{\tt arXiv:2107.14779}}].

\bibitem{Cangemi:2022abk}
L.~Cangemi and P.~Pichini, {\it {Classical limit of higher-spin string
  amplitudes}},  {\em JHEP} {\bf 06} (2023) 167,
  [\href{http://arxiv.org/abs/2207.03947}{{\tt arXiv:2207.03947}}].

\bibitem{Cangemi:2022bew}
L.~Cangemi, M.~Chiodaroli, H.~Johansson, A.~Ochirov, P.~Pichini, and
  E.~Skvortsov, {\it {Kerr Black Holes Enjoy Massive Higher-Spin Gauge
  Symmetry}},  \href{http://arxiv.org/abs/2212.06120}{{\tt arXiv:2212.06120}}.

\bibitem{Haddad:2021znf}
K.~Haddad, {\it {Exponentiation of the leading eikonal phase with spin}},  {\em
  Phys. Rev. D} {\bf 105} (2022), no.~2 026004,
  [\href{http://arxiv.org/abs/2109.04427}{{\tt arXiv:2109.04427}}].

\bibitem{Aoude:2022trd}
R.~Aoude, K.~Haddad, and A.~Helset, {\it {Searching for Kerr in the 2PM
  amplitude}},  {\em JHEP} {\bf 07} (2022) 072,
  [\href{http://arxiv.org/abs/2203.06197}{{\tt arXiv:2203.06197}}].

\bibitem{Menezes:2022tcs}
G.~Menezes and M.~Sergola, {\it {NLO deflections for spinning particles and
  Kerr black holes}},  {\em JHEP} {\bf 10} (2022) 105,
  [\href{http://arxiv.org/abs/2205.11701}{{\tt arXiv:2205.11701}}].

\bibitem{Bern:2022kto}
Z.~Bern, D.~Kosmopoulos, A.~Luna, R.~Roiban, and F.~Teng, {\it {Binary Dynamics
  through the Fifth Power of Spin at $O(G^2)$}},  {\em Phys. Rev. Lett.} {\bf
  130} (2023), no.~20 201402, [\href{http://arxiv.org/abs/2203.06202}{{\tt
  arXiv:2203.06202}}].

\bibitem{Alessio:2022kwv}
F.~Alessio and P.~Di~Vecchia, {\it {Radiation reaction for spinning black-hole
  scattering}},  {\em Phys. Lett. B} {\bf 832} (2022) 137258,
  [\href{http://arxiv.org/abs/2203.13272}{{\tt arXiv:2203.13272}}].

\bibitem{Alessio:2023kgf}
F.~Alessio, {\it {Kerr binary dynamics from minimal coupling and double copy}},
   \href{http://arxiv.org/abs/2303.12784}{{\tt arXiv:2303.12784}}.

\bibitem{Bjerrum-Bohr:2023jau}
N.~E.~J. Bjerrum-Bohr, G.~Chen, and M.~Skowronek, {\it {Classical spin
  gravitational Compton scattering}},  {\em JHEP} {\bf 06} (2023) 170,
  [\href{http://arxiv.org/abs/2302.00498}{{\tt arXiv:2302.00498}}].

\bibitem{Damgaard:2022jem}
P.~H. Damgaard, J.~Hoogeveen, A.~Luna, and J.~Vines, {\it {Scattering angles in
  Kerr metrics}},  {\em Phys. Rev. D} {\bf 106} (2022), no.~12 124030,
  [\href{http://arxiv.org/abs/2208.11028}{{\tt arXiv:2208.11028}}].

\bibitem{Haddad:2023ylx}
K.~Haddad, {\it {Recursion in the classical limit and the neutron-star Compton
  amplitude}},  {\em JHEP} {\bf 05} (2023) 177,
  [\href{http://arxiv.org/abs/2303.02624}{{\tt arXiv:2303.02624}}].

\bibitem{Aoude:2023vdk}
R.~Aoude, K.~Haddad, and A.~Helset, {\it {Classical gravitational scattering
  amplitude at O(G2S1\ensuremath{\infty}S2\ensuremath{\infty})}},  {\em Phys.
  Rev. D} {\bf 108} (2023), no.~2 024050,
  [\href{http://arxiv.org/abs/2304.13740}{{\tt arXiv:2304.13740}}].

\bibitem{Jakobsen:2023ndj}
G.~U. Jakobsen, G.~Mogull, J.~Plefka, B.~Sauer, and Y.~Xu, {\it {Conservative
  Scattering of Spinning Black Holes at Fourth Post-Minkowskian Order}},  {\em
  Phys. Rev. Lett.} {\bf 131} (2023), no.~15 151401,
  [\href{http://arxiv.org/abs/2306.01714}{{\tt arXiv:2306.01714}}].

\bibitem{Jakobsen:2023hig}
G.~U. Jakobsen, G.~Mogull, J.~Plefka, and B.~Sauer, {\it {Dissipative
  scattering of spinning black holes at fourth post-Minkowskian order}},
  \href{http://arxiv.org/abs/2308.11514}{{\tt arXiv:2308.11514}}.

\bibitem{Heissenberg:2023uvo}
C.~Heissenberg, {\it {Angular Momentum Loss Due to Spin-Orbit Effects in the
  Post-Minkowskian Expansion}},  \href{http://arxiv.org/abs/2308.11470}{{\tt
  arXiv:2308.11470}}.

\bibitem{Bianchi:2023lrg}
M.~Bianchi, C.~Gambino, and F.~Riccioni, {\it {A Rutherford-like formula for
  scattering off Kerr-Newman BHs and subleading corrections}},  {\em JHEP} {\bf
  08} (2023) 188, [\href{http://arxiv.org/abs/2306.08969}{{\tt
  arXiv:2306.08969}}].

\bibitem{Westpfahl:1979gu}
K.~Westpfahl and M.~Goller, {\it {GRAVITATIONAL SCATTERING OF TWO RELATIVISTIC
  PARTICLES IN POSTLINEAR APPROXIMATION}},  {\em Lett. Nuovo Cim.} {\bf 26}
  (1979) 573--576.

\bibitem{Damour:1990jh}
T.~Damour and G.~Schaefer, {\it {Redefinition of position variables and the
  reduction of higher order Lagrangians}},  {\em J. Math. Phys.} {\bf 32}
  (1991) 127--134.

\bibitem{Buonanno:2000qq}
A.~Buonanno, {\it {Reduction of the two-body dynamics to a one-body description
  in classical electrodynamics}},  {\em Phys. Rev. D} {\bf 62} (2000) 104022,
  [\href{http://arxiv.org/abs/hep-th/0004042}{{\tt hep-th/0004042}}].

\bibitem{Saketh:2021sri}
M.~V.~S. Saketh, J.~Vines, J.~Steinhoff, and A.~Buonanno, {\it {Conservative
  and radiative dynamics in classical relativistic scattering and bound
  systems}},  {\em Phys. Rev. Res.} {\bf 4} (2022), no.~1 013127,
  [\href{http://arxiv.org/abs/2109.05994}{{\tt arXiv:2109.05994}}].

\bibitem{Bern:2021xze}
Z.~Bern, J.~P. Gatica, E.~Herrmann, A.~Luna, and M.~Zeng, {\it {Scalar QED as a
  toy model for higher-order effects in classical gravitational scattering}},
  {\em JHEP} {\bf 08} (2022) 131, [\href{http://arxiv.org/abs/2112.12243}{{\tt
  arXiv:2112.12243}}].

\bibitem{Bern:2023ccb}
Z.~Bern, E.~Herrmann, R.~Roiban, M.~S. Ruf, A.~V. Smirnov, V.~A. Smirnov, and
  M.~Zeng, {\it {Conservative binary dynamics at order $O(\alpha^5)$ in
  electrodynamics}},  \href{http://arxiv.org/abs/2305.08981}{{\tt
  arXiv:2305.08981}}.

\bibitem{DiVecchia:2021bdo}
P.~Di~Vecchia, C.~Heissenberg, R.~Russo, and G.~Veneziano, {\it {The eikonal
  approach to gravitational scattering and radiation at $ \mathcal{O}
  $(G$^{3}$)}},  {\em JHEP} {\bf 07} (2021) 169,
  [\href{http://arxiv.org/abs/2104.03256}{{\tt arXiv:2104.03256}}].

\bibitem{DiVecchia:2022piu}
P.~Di~Vecchia, C.~Heissenberg, R.~Russo, and G.~Veneziano, {\it {Classical
  gravitational observables from the Eikonal operator}},  {\em Phys. Lett. B}
  {\bf 843} (2023) 138049, [\href{http://arxiv.org/abs/2210.12118}{{\tt
  arXiv:2210.12118}}].

\bibitem{DiVecchia:2023frv}
P.~Di~Vecchia, C.~Heissenberg, R.~Russo, and G.~Veneziano, {\it {The
  gravitational eikonal: from particle, string and brane collisions to
  black-hole encounters}},  \href{http://arxiv.org/abs/2306.16488}{{\tt
  arXiv:2306.16488}}.

\bibitem{FebresCordero:2022jts}
F.~Febres~Cordero, M.~Kraus, G.~Lin, M.~S. Ruf, and M.~Zeng, {\it {Conservative
  Binary Dynamics with a Spinning Black Hole at O(G3) from Scattering
  Amplitudes}},  {\em Phys. Rev. Lett.} {\bf 130} (2023), no.~2 021601,
  [\href{http://arxiv.org/abs/2205.07357}{{\tt arXiv:2205.07357}}].

\bibitem{Jakobsen:2022fcj}
G.~U. Jakobsen and G.~Mogull, {\it {Conservative and Radiative Dynamics of
  Spinning Bodies at Third Post-Minkowskian Order Using Worldline Quantum Field
  Theory}},  {\em Phys. Rev. Lett.} {\bf 128} (2022), no.~14 141102,
  [\href{http://arxiv.org/abs/2201.07778}{{\tt arXiv:2201.07778}}].

\bibitem{Jakobsen:2022zsx}
G.~U. Jakobsen and G.~Mogull, {\it {Linear response, Hamiltonian, and radiative
  spinning two-body dynamics}},  {\em Phys. Rev. D} {\bf 107} (2023), no.~4
  044033, [\href{http://arxiv.org/abs/2210.06451}{{\tt arXiv:2210.06451}}].

\bibitem{Bini:2023mdz}
D.~Bini, A.~Geralico, and P.~Rettegno, {\it {Spin-orbit contribution to
  radiative losses for spinning binaries with aligned spins}},  {\em Phys. Rev.
  D} {\bf 108} (2023), no.~6 064049,
  [\href{http://arxiv.org/abs/2307.12670}{{\tt arXiv:2307.12670}}].

\bibitem{Aoude:2023dui}
R.~Aoude, K.~Haddad, C.~Heissenberg, and A.~Helset, {\it {Leading-order
  gravitational radiation to all spin orders}},
  \href{http://arxiv.org/abs/2310.05832}{{\tt arXiv:2310.05832}}.

\bibitem{Brandhuber:2023hhl}
A.~Brandhuber, G.~R. Brown, G.~Chen, J.~Gowdy, and G.~Travaglini, {\it
  {Resummed spinning waveforms from five-point amplitudes}},
  \href{http://arxiv.org/abs/2310.04405}{{\tt arXiv:2310.04405}}.

\bibitem{DeAngelis:2023lvf}
S.~De~Angelis, R.~Gonzo, and P.~P. Novichkov, {\it {Spinning waveforms from
  KMOC at leading order}},  \href{http://arxiv.org/abs/2309.17429}{{\tt
  arXiv:2309.17429}}.

\bibitem{Singh:1974qz}
L.~P.~S. Singh and C.~R. Hagen, {\it {Lagrangian formulation for arbitrary
  spin. 1. The boson case}},  {\em Phys. Rev. D} {\bf 9} (1974) 898--909.

\bibitem{Lindwasser:2023dcv}
L.~W. Lindwasser, {\it {Consistent actions for massive particles interacting
  with electromagnetism and gravity}},
  \href{http://arxiv.org/abs/2309.03901}{{\tt arXiv:2309.03901}}.

\bibitem{Lindwasser:2023zwo}
L.~W. Lindwasser, {\it {Covariant actions and propagators for all spins,
  masses, and dimensions}},  \href{http://arxiv.org/abs/2307.11750}{{\tt
  arXiv:2307.11750}}.

\bibitem{arkanihamed2021scattering}
N.~Arkani-Hamed, T.-C. Huang, and Y.~tin Huang, {\it Scattering amplitudes for
  all masses and spins},  2021.

\bibitem{Johansson:2019dnu}
H.~Johansson and A.~Ochirov, {\it {Double copy for massive quantum particles
  with spin}},  {\em JHEP} {\bf 09} (2019) 040,
  [\href{http://arxiv.org/abs/1906.12292}{{\tt arXiv:1906.12292}}].

\bibitem{Herrmann:2021tct}
E.~Herrmann, J.~Parra-Martinez, M.~S. Ruf, and M.~Zeng, {\it {Radiative
  classical gravitational observables at $ \mathcal{O} $(G$^{3}$) from
  scattering amplitudes}},  {\em JHEP} {\bf 10} (2021) 148,
  [\href{http://arxiv.org/abs/2104.03957}{{\tt arXiv:2104.03957}}].

\bibitem{Steinhoff:2014kwa}
J.~Steinhoff, {\it {Spin and quadrupole contributions to the motion of
  astrophysical binaries}},  {\em Fund. Theor. Phys.} {\bf 179} (2015)
  615--649, [\href{http://arxiv.org/abs/1412.3251}{{\tt arXiv:1412.3251}}].

\bibitem{Steinhoff:2015ksa}
J.~Steinhoff, {\it {Spin gauge symmetry in the action principle for classical
  relativistic particles}},  \href{http://arxiv.org/abs/1501.04951}{{\tt
  arXiv:1501.04951}}.

\bibitem{Vines:2016unv}
J.~Vines, D.~Kunst, J.~Steinhoff, and T.~Hinderer, {\it {Canonical Hamiltonian
  for an extended test body in curved spacetime: To quadratic order in spin}},
  {\em Phys. Rev. D} {\bf 93} (2016), no.~10 103008,
  [\href{http://arxiv.org/abs/1601.07529}{{\tt arXiv:1601.07529}}]. [Erratum:
  Phys.Rev.D 104, 029902 (2021)].

\bibitem{citeycite}
J.~Ehlers and E.~Rudolph, {\it Dynamics of extended bodies in general
  relativity center-of-mass description and quasirigidity},  {\em General
  Relativity and Gravitation} {\bf 8} (1977), no.~3 197--217.

\bibitem{Parra-Martinez:2020dzs}
J.~Parra-Martinez, M.~S. Ruf, and M.~Zeng, {\it {Extremal black hole scattering
  at $\mathcal{O}(G^3)$: graviton dominance, eikonal exponentiation, and
  differential equations}},  {\em JHEP} {\bf 11} (2020) 023,
  [\href{http://arxiv.org/abs/2005.04236}{{\tt arXiv:2005.04236}}].

\bibitem{Forde:2007mi}
D.~Forde, {\it {Direct extraction of one-loop integral coefficients}},  {\em
  Phys. Rev. D} {\bf 75} (2007) 125019,
  [\href{http://arxiv.org/abs/0704.1835}{{\tt arXiv:0704.1835}}].

\bibitem{Herderschee:2023fxh}
A.~Herderschee, R.~Roiban, and F.~Teng, {\it {The sub-leading scattering
  waveform from amplitudes}},  {\em JHEP} {\bf 06} (2023) 004,
  [\href{http://arxiv.org/abs/2303.06112}{{\tt arXiv:2303.06112}}].

\bibitem{Adamo:2022qci}
T.~Adamo, A.~Cristofoli, A.~Ilderton, and S.~Klisch, {\it {All Order
  Gravitational Waveforms from Scattering Amplitudes}},  {\em Phys. Rev. Lett.}
  {\bf 131} (2023), no.~1 011601, [\href{http://arxiv.org/abs/2210.04696}{{\tt
  arXiv:2210.04696}}].

\bibitem{Jones:2023ugm}
C.~R.~T. Jones and M.~S. Ruf, {\it {Absorptive Effects and Classical Black Hole
  Scattering}},  \href{http://arxiv.org/abs/2310.00069}{{\tt
  arXiv:2310.00069}}.

\bibitem{Jones:2022aji}
C.~R.~T. Jones and M.~Solon, {\it {Scattering amplitudes and N-body
  post-Minkowskian Hamiltonians in general relativity and beyond}},  {\em JHEP}
  {\bf 02} (2023) 105, [\href{http://arxiv.org/abs/2208.02281}{{\tt
  arXiv:2208.02281}}].

\bibitem{Weinberg:1995mt}
S.~Weinberg, {\em {The Quantum theory of fields. Vol. 1: Foundations}}.
\newblock Cambridge University Press, 6, 2005.

\end{thebibliography}\endgroup

\end{document}